\documentclass[sigconf, nonacm]{acmart}
\usepackage[utf8]{inputenc}
\usepackage{graphicx}
\usepackage{soul}
\usepackage{enumitem}
\usepackage{subcaption}

\newcommand{\hide}[1]{}
 \def\@copyrightspace{\relax}
\newcommand{\xhdr}[1]{\vspace{0.3mm}\noindent{{\bf #1.}}}
\newcommand{\xhdrNoPeriod}[1]{\vspace{0.3mm}\noindent{{\bf #1}}}

\newcommand{\numchannels}{349 }
\newcommand{\numvideos}{330,925 }
\newcommand{\numcommentsapprox}{72M }
\newcommand{\numcommentsapproxplus}{72M+ }
\newcommand{\numcomments}{72,069,878 }
\newcommand{\numrecsapprox}{2M }
\newcommand{\numrecschanapprox}{10K }

\AtBeginDocument{%
  \providecommand\BibTeX{{%
    \normalfont B\kern-0.5em{\scshape i\kern-0.25em b}\kern-0.8em\TeX}}}
\setcopyright{none}

\copyrightyear{2020}
\acmYear{2020}
\setcopyright{acmlicensed}
\acmConference[FAT* '20]{Conference on Fairness, Accountability, and Transparency}{January 27--30, 2020}{Barcelona, Spain}
\acmBooktitle{Conference on Fairness, Accountability, and Transparency (FAT* '20), January 27--30, 2020, Barcelona, Spain}
\acmPrice{15.00}
\acmDOI{10.1145/3351095.3372879}
\acmISBN{978-1-4503-6936-7/20/02}

\begin{document}

\makeatletter
\def\@copyrightspace{\relax}
\makeatother

\title{Auditing Radicalization Pathways on YouTube}
% \author{Anonymous
% }
% \email{ }
% \affiliation{%
%   \institution{ }
% }
\author{Manoel Horta Ribeiro}
\authornote{Work done mostly while at UFMG.}
\affiliation{%
  \institution{EPFL}
}
\email{manoel.hortaribeiro@epfl.ch}

\author{Raphael Ottoni}
\affiliation{%
  \institution{UFMG}
}
\email{rapha@dcc.ufmg.br}

\author{Robert West}
\affiliation{%
  \institution{EPFL}
}
\email{robert.west@epfl.ch}

\author{Virgílio A. F. Almeida}
\affiliation{%
  \institution{UFMG, Berkman Klein Center}
}
\email{virgilio@dcc.ufmg.br}

\author{Wagner Meira Jr.}
\affiliation{%
  \institution{UFMG}
}
\email{meira@dcc.ufmg.br}

\renewcommand{\shortauthors}{Horta Ribeiro et al.}

\begin{abstract}
Non-profits, as well as the media, have hypothesized the existence of a radicalization pipeline on YouTube, claiming that users systematically progress towards more extreme content on the platform.
Yet, there is to date no substantial quantitative evidence of this alleged pipeline.
To close this gap, we conduct a large-scale audit of user radicalization on YouTube.
We analyze \numvideos videos posted on \numchannels channels, which we broadly classified into four types: Media, the Alt-lite, the Intellectual Dark Web (I.D.W.), and the Alt-right.
According to the aforementioned radicalization hypothesis, channels in the I.D.W. and the Alt-lite serve as gateways to fringe far-right ideology, here represented by Alt-right channels.
Processing \numcommentsapproxplus comments, we show that the three channel types indeed increasingly share the same user base; 
that users consistently migrate from milder to more extreme content; 
and that a large percentage of users who consume Alt-right content now consumed Alt-lite and I.D.W. content in the past.
We also probe YouTube's recommendation algorithm, looking at more than \numrecsapprox video and channel recommendations between May/July 2019.
We find that Alt-lite content is easily reachable from I.D.W. channels, while Alt-right videos are reachable only through channel recommendations.
Overall, we paint a comprehensive picture of user radicalization on YouTube.
\end{abstract}




\maketitle

\newpage
\section{Introduction}
\label{sec:001}

Video channels that discuss social, political and cultural subjects have flourished on YouTube.
Frequently, the videos posted in such channels focus on highly controversial topics such as race, gender, and religion.
The users who create and post such videos span a wide spectrum of political orientation, from prolific podcast hosts like Joe Rogan to outspoken advocates of white supremacy like Richard Spencer.
These individuals not only share the same platform but often publicly engage in debates and conversations with each other on the website~\cite{lewisAlternativeInfluenceBroadcasting2018}.
This way, even distant personalities can be linked in chains of pairwise co-appearances.
For instance, Joe Rogan interviewed YouTuber Carl Benjamin~\cite{powerfuljreSargonAkkadJoeYouTube}, who debated with white supremacist Richard Spencer~\cite{andywarskiRichardSpencerStyxYoutube}.

According to Lewis~\cite{lewisAlternativeInfluenceBroadcasting2018}, this proximity may create ``radicalization pathways'' for audience members and content creators.
Examples of these journeys are plenty, including content creator Roosh V's trajectory from pick-up artist to Alt-right supporter~\cite{kutnerRooshJourneyPickup2016, rooshvNotDisavowRichard2016} and Caleb Cain's testimony of his YouTube-driven radicalization~\cite{rooseMakingYouTubeRadical2019}.

The claim that there is a ``radicalization pipeline'' on YouTube should be considered in the  context of decreasing trust in mainstream media and increasing influence of social networks.
Across the globe, individuals are skeptical of traditional media vehicles and growingly consume news and opinion content on social media~\cite{nicReutersInstituteDigital2018, ingramMostAmericansSay2018}. 
In this setting, recent research has shown that fringe websites (\emph{e.g.}, \textit{4chan}) and subreddits (\emph{e.g.},  \textit{/r/TheDonald}) have great influence over which memes~\cite{zannettouOriginsMemesMeans2018} and news~\cite{zannettouWebCentipedeUnderstanding2017} are shared in large social networks, such as Twitter. 
YouTube is extremely popular, especially among children and teenagers~\cite{andersonTeensSocialMedia2018}, and if the streaming website is actually radicalizing individuals this could push fringe ideologies like white supremacy further into the mainstream~\cite{tufekciOpinionYouTubeGreat2018}.

A key issue in dealing with topics like radicalization and hate speech is the lack of agreement over what is ``hateful'' or ``extreme''~\cite{sellarsDefiningHateSpeech2016}.
A workaround is to perform analyses based on \textit{communities}, large sets of loosely associated content creators (here represented by their YouTube channels).
For the purpose of this work, we consider three ``communities'' that have been associated with user radicalization~\cite{lewisAlternativeInfluenceBroadcasting2018, weissOpinionMeetRenegades2018, rooseMakingYouTubeRadical2019} and that differ in the extremity of their content:
the ``Intellectual Dark Web'' (I.D.W.), the ``Alt-lite'' and the ``Alt-right''.
While users in the I.D.W. discuss controversial subjects like race and I.Q.~\cite{weissOpinionMeetRenegades2018} without necessarily endorsing extreme views, members of the Alt-right sponsor fringe ideas like that of a white ethnostate~\cite{hankesAltRightKillingPeople}. 
Somewhere in the middle, individuals of the Alt-lite deny to embrace white supremacist ideology, although they frequently flirt with concepts associated with it (\emph{e.g}., the ``Great Replacement'', globalist conspiracies).

\vspace{2mm}

\xhdr{Present work} 
In this paper, we audit whether users are indeed becoming radicalized on YouTube and whether the recommendation algorithms contribute towards this radicalization.
We do so by examining three prominent communities: 
the Intellectual Dark Web, the Alt-lite and the Alt-right.
More specifically, considering Alt-right content as a proxy for extreme content, we ask:

\begin{itemize}
\item[\textbf{RQ1}] How have these channels grown on YouTube in the last decade?
\item[\textbf{RQ2}] To which extent do users systematically gravitate towards more extreme content?
\item[\textbf{RQ3}] Do algorithmic recommendations steer users towards more extreme content? 
\end{itemize}

\noindent
We develop a data collection process where we 
\textit{(i)}~acquire a large pool of relevant channels from these communities;
\textit{(ii)}~collect metadata and comments for each of the videos in the channels; 
\textit{(iii)}~annotate channels as belonging to several different communities; and
\textit{(iv)}~collect YouTube video and channel recommendations.
We also collect traditional and alternative media channels for additional comparisons.
We use these as a sanity check to capture the growth of other content on YouTube, rather than trying to obtain similar users in other channels.
These efforts resulted in a dataset with more than \numcommentsapprox comments in \numvideos videos of \numchannels channels and with more than \numrecsapprox video and \numrecschanapprox channel recommendations.
Importantly, our recommendations do not account for personalization.
We analyze this large dataset extensively:

\begin{itemize}

\item 
We look at the growth of the I.D.W., the Alt-lite and the Alt-right throughout the last decade in terms of videos, likes and views, finding a steep rise in activity and engagement in the communities of interest when compared with the media channels.
Moreover, comments per view seem to be particularly high in more extreme content, reaching near to 1 comment for every 5 views in Alt-right channels in 2018 (Sec.~\ref{sec:004}).

\item 
We inspect the intersection of commenting users across the communities, finding they increasingly share the same user base.
Analyzing the overlap between the sets of commenting users, we find that approximately half of the users who commented on Alt-right channels in 2018 also comment on Alt-lite and on I.D.W. channels (Sec.~\ref{sec:005}).

\item 
We also find that the intersection is not only growing due to new users but that there is significant user migration among the communities being studied. 
Users that initially comment only on content from the I.D.W. or the Alt-lite throughout the years consistently start to comment on Alt-right content. These users are a significant fraction of the Alt-right commenting user base. 
This effect is much stronger than for the large traditional and alternative media channels we collected (Sec.~\ref{sec:006}).

\item 
Lastly, we take a look at the impact of YouTube's recommendation algorithms, running simulations on recommendation graphs. 
Our analyses show that, particularly through the channel recommender system, Alt-lite channels are easily discovered from I.D.W. channels, and that Alt-right channels may be reached from the two other communities (Sec.~\ref{sec:007}).

\end{itemize}
\vspace{2.5mm}

This is, to our best knowledge, the first large scale quantitative audit of user radicalization on YouTube. 
We find strong evidence for radicalization among YouTube users, and that YouTube's recommender system enables Alt-right channels to be discovered, even in a scenario without personalization.
We discuss our findings and our limitations further in Sec.~\ref{sec:008}. 
We argue that commenting users are a good enough proxy to measure the user radicalization, as more extreme content seems to beget more comments.
Moreover, regardless of the degree of influence of the recommender system in the process of radicalizing users, there is significant evidence that users are reaching content sponsoring fringe ideologies from the Alt-lite and the Intellectual Dark Web.

\section{Background}
\label{sec:002}
\xhdr{Contrarian communities} 
We discuss three of YouTube's prominent communities: the Intellectual Dark Web, the Alt-lite and the Alt-right.
We argue that all of them are \textit{contrarians}, in the sense that they often oppose mainstream views or attitudes. 
According to Nagle, these communities flourished in the wave of ``anti-PC'' culture of the 2010s, where social-political movements (e.g. the transgender rights movement, the anti-sexual assault movement) were  portrayed as hysterical, and their claims, as absurd~\cite{nagleKillAllNormies2017}.

According to the Anti Defamation League~\cite{AltRightAlt2019ADL}, the Alt-Right is a loose segment of the white supremacist movement consisting of individuals who reject mainstream conservatism in favor of politics that embrace racist, anti-Semitic and white supremacist ideology. 
The Alt-right skews younger than other far-right groups, and has a big online presence, particularly on fringe web sites like \emph{4chan}, \emph{8chan} and certain corners of \emph{Reddit}~\cite{AltRightAnti-DefamationLeague}. 

The term Alt-lite  was created to differentiate right-wing activists who deny  embracing white supremacist ideology.
Atkison argues that the Unite the Rally in Charlottesville was deeply related to this change, as participants of the rally revealed the movement's white supremacist leanings and affiliations~\cite{atkinsonCharlottesvilleAltrightTurning2018}.
Alt-right writer and white supremacist Greg Johnson~\cite{AltRightAlt2019ADL} describes the difference between Alt-right and Alt-lite by the origin of its nationalism: \textit{"The Alt-lite is defined by civic nationalism as opposed to racial nationalism, which is a defining characteristic of the Alt-right"}. 
This distinction was also highlighted in~\cite{marantzAltRightBrandingWar2017}. 
Yet it is important to point out that the line between the Alt-right and the Alt-lite is blurry~\cite{AltRightAlt2019ADL}, as many Alt-liters are accused of dog-whistling: attenuating their real beliefs to appeal to a more general public and to prevent getting banned~\cite{lopezg.StefanMolyneuxMAGA2019, joelkellyLaurenSouthernAltright2017}. 
To address this problem, in this paper we take a conservative approach to our labeling, naming only the most extreme content creators as Alt-right. 

The ``Intellectual Dark Web'' (I.D.W.) is a term coined by Eric Weinstein to refer to a group of academics and podcast hosts~\cite{weissOpinionMeetRenegades2018}.
The neologism was popularized  in a New York Times opinion article~\cite{weissOpinionMeetRenegades2018}, where it is used to describe ``iconoclastic thinkers, academic renegades and media personalities who are having a rolling conversation about all sorts of subjects, [\dots] touching on controversial issues such as abortion, biological differences between men and women, identity politics, religion, immigration, etc.''

The group described in the NYT piece includes, among others, Sam Harris, Jordan Peterson, Ben Shapiro, Dave Rubin, and Joe Rogan, and also mentions a website with an unofficial list of members~\cite{anonymousIntellectualDarkWeb}. 
Members of the so-called I.D.W. have been accused of espousing politically incorrect ideas \cite{beydounUSLiberalIslamophobia2018,JordanPetersonTransgender2017SpectatorLife,foderaroAlexandriaOcasioCortezLikens2018}. 
Moreover, a recent report by the Data \& Society Research Institute has claimed these channels are ``pathways to radicalization''~\cite{lewisAlternativeInfluenceBroadcasting2018}, acting as entry points to more radical channels, such as those in Alt-right.  
Broadly, members of this loosely defined movement see these criticisms as a consequence of discussing controversial subjects~\cite{weissOpinionMeetRenegades2018}, and some have explicitly dismissed the report~\cite{therubinreportEricWeinsteinFuture2018}.
Similarly to what happens between Alt-right and Alt-lite, there are also blurry lines between the I.D.W. and the Alt-lite, especially for non-core members, such as those listed on the aforementioned website~\cite{anonymousIntellectualDarkWeb}.
To break ties, we label borderline cases as Alt-lite.

\xhdr{Radicalization} 
We consider the definition given by McCauley and Moskalenko \cite{mccauleyMechanismsPoliticalRadicalization2008}:
(``Functionally, political radicalization is increased preparation for and commitment to intergroup conflict. 
Descriptively, radicalization means change in beliefs, feelings, and behaviors in directions that increasingly justify intergroup violence and demand sacrifice in defense of the ingroup.'')
and use increased consumption of Alt-right content as a proxy for radicalization. 
This is reasonable since the Alt-right's rhetoric has been invoked by the perpetrators of some recent terrorist attacks (\textit{e.g.} the Christchurch mosque shooting~\cite{mannEmperorCottrellAccused2019ABCNews}), and since it champions ideas promoting intergroup conflict (\textit{e.g.} a white ethnostate~\cite{hankesAltRightKillingPeople}). 
Our conservative strategy when labeling channels is of particular importance here: Alt-right channels are closely related to these ideas, while the Alt-lite/I.D.W. are given the benefit of doubt.

\xhdr{Auditing the web} 
As algorithms play an ever-larger role in our lives, it is increasingly important for researchers and society at large to reverse engineer algorithms' input-output relationships~\cite{diakopoulosAlgorithmicAccountabilityReporting2014}. 
Previous large scale algorithmic auditing include measuring discrimination on AirBnB~\cite{edelmanDigitalDiscriminationCase2014}, personalization on web search~\cite{hannakMeasuringPersonalizationWeb2013} and price discrimination on e-commerce web sites~\cite{hannakMeasuringPriceDiscrimination2014}.
We argue this work is an audit in the sense that it measures a troublesome phenomenon (user radicalization) in a content-sharing social environment heavily influenced by algorithms (YouTube). 
Unfortunately, it is not possible to obtain the entire history of YouTube recommendation, so we must limit the algorithmic analyses to a time slice of a constantly changing black-box. 
Although comments may give us insight into the past, it is challenging to tease apart the influence of the algorithm in previous times. 
Another limitation of our auditing is that we do not account for user personalization. 
Despite these flaws, we argue that: \textit{(i)} our analyses provide answers to important questions related with impactful societal processes that are allegedly happening in YouTube (regardless of the impact of the recommender system), and \textit{(ii)} our framework for auditing user radicalization can be replicated through time and expanded to handle personalization.

\xhdr{Previous research from/on YouTube} 
Previous work by Google sheds light into some of the high-level technicalities of YouTube's recommender system~\cite{covingtonDeepNeuralNetworks2016, davidsonYouTubeVideoRecommendation2010}.
Their latest paper indicates they  use embeddings for video searches and video histories as inputs for a dense neural network~\cite{davidsonYouTubeVideoRecommendation2010}. 
There also exists a large body of work studying violent~\cite{giannakopoulosMultimodalApproachViolence2010}, hateful or  extremist~\cite{surekaMiningYouTubeDiscover2010, agarwalFocusedCrawlerMining2014} and disturbing content~\cite{papadamouDisturbedYouTubeKids2019} on the platform.
Much of the existing work focuses on creating detection algorithms for these types of content using features of the comments, the commenting users and the videos~\cite{agarwalFocusedCrawlerMining2014, giannakopoulosMultimodalApproachViolence2010}.
Sureka et al.~\cite{surekaMiningYouTubeDiscover2010} use a seed-expanding methodology to track extremist user communities, which yielded high precision in including relevant users. 
This is somewhat analogous to what we do, although we use YouTube's recommender system while they use user friends, subscriptions and favorites. 
Ottoni et al. perform an in-depth textual analysis of 23 channels (13 broadly defined as Alt-right), finding significantly different topics across the two groups~\cite{ottoniAnalyzingRightwingYouTube2018}.
O'Callegan et al.~\cite{ocallaghanWhiteRabbitHole2015} simulate a recommender system with channels tweeted in an extreme right dataset. They show that a simple non-negative matrix factorization metadata-based recommender system would cluster extreme right topics together.

\section{Data Collection}
\label{sec:003}
We are interested in three communities on YouTube: the I.D.W., the Alt-lite, and the Alt-right.
Identifying such communities and the channels which belong to them is no easy task: 
 the membership of channels to these communities is volatile and fuzzy,
 and there is disagreement between how members of these communities view themselves, and how they are considered by scholars and the media.
These particularities make our challenge multi-faceted: on one hand, we want to study user radicalization, and determine, for example, if users who start watching videos by communities like the I.D.W. eventually go on to consume Alt-right content. 
On the other, there is often no clear agreement on who belongs to which community.

Due to these nuances, we devise a careful methodology to 
\textbf{(a)}~collect a large pool of relevant channels;
\textbf{(b)}~collect data and the recommendations given by YouTube for these channels; 
\textbf{(c)}~manually labeling these channels according to the communities of interest.

\begin{table*}[t]
\caption{Top $16$ YouTube channels with the most views per each community and for media channels.}
\small
\begin{tabular}{lllllllll}
\toprule
{} &                   Alt-right & Views &            Alt-lite & View  &  Intellectual Dark Web & Views   &                     Media & Views    \\
\midrule
1  &                James Allsup &        62M &       StevenCrowder &        727M &            PowerfulJRE &           1B &                         vox &            1B \\
2  &         Black Pigeon Speaks &        50M &         Rebel Media &        405M &              JRE Clips &         717M &                 gq magazine &            1B \\
3  &          ThuleanPerspective &        45M &  Paul Joseph Watson &        356M &       PragerUniversity &         635M &                   vice news &            1B \\
4  &                  Red Ice TV &        42M &            MarkDice &        334M &         The Daily Wire &         247M &              wired magazine &            1B \\
5  &              The Golden One &        12M &    SargonofAkkad100 &        258M &       The Rubin Report &         206M &                 vanity fair &          639M \\
6  &                 AmRenVideos &         9M &     Stefan Molyneux &        193M &               ReasonTV &         138M &                   the verge &          636M \\
7  &    NeatoBurrito Productions &         7M &        hOrnsticles3 &        145M &   JordanPetersonVideos &          90M &            glamour magazine &          620M \\
8  &              The Last Stand &         7M &                MILO &        133M &  Bite-sized Philosophy &          62M &            business insider &          523M \\
9  &              MillennialWoes &         6M &  Styxhexenhammer666 &        132M &          Owen Benjamin &          35M &  huffington post &          329M \\
10 &                Mark Collett &         6M &       OneTruth4Life &        112M &       AgatanFoundation &          33M &           today i found out &          328M \\
11 &           AustralianRealist &         5M &         No Bullshit &        104M &        Essential Truth &          32M &                    cbc news &          324M \\
12 &       Jean-François Gariépy &         5M &          SJWCentral &         90M &            Ben Shapiro &          30M &                the guardian &          300M \\
13 &          Prince of Zimbabwe &         5M &   Computing Forever &         87M &                  YAFTV &          30M &             people magazine &          287M \\
14 &  The Alternative Hypothesis &         5M &        The Thinkery &         86M &         joerogandotnet &          25M &                   big think &          258M \\
15 &               Matthew North &         4M &             Bearing &         81M &        TheArchangel911 &          24M &                cosmopolitan &          256M \\
16 &               Faith J Goldy &         4M &       RobinHoodUKIP &         64M &         Clash of Ideas &          24M &                 global news &          252M \\
\bottomrule
\end{tabular}

\label{tab:tablechannels}
\end{table*}

\xhdrNoPeriod{(a)} For each community, we create a pool of channels as follows. 
We refer to channels obtained in the $i$-th step as \textit{Type $i$} channels.

\begin{enumerate}[leftmargin = 0.6cm]

\item 
We choose a set of \textit{seed channels}.
Seeds were extracted from the I.D.W. unofficial website~\cite{anonymousIntellectualDarkWeb}, Anti Defamation League's report on the Alt-lite/the Alt-right~\cite{AltRightAlt2019ADL} and Data \& Society's report on YouTube Radicalization~\cite{lewisAlternativeInfluenceBroadcasting2018}. 
We pick popular channels that are representative of the community we are interested in. 
Each seed was independently annotated two times and discarded in case there was any disagreement. 
We further detail the annotation process later in this section.

\item 
We choose a set of \textit{keywords} related to the sub-communities. 
For each keyword, we use YouTube's search functionality and consider the first $200$ results in English.
We then add channels that broadly relate in topic to the community in question.
For example, for the Alt-right, keywords included both terms associated with their narratives, such as \textit{The Jewish Question} and \textit{White Genocide,} as well as the names or nicknames of famous Alt-righters, such as \textit{weev} and \textit{Christopher Cantwell.}

\item 
We iteratively search the related and featured channels collected in steps (1) and (2), adding relevant channels (as defined in 2). 
Note that these are two ways channel can link to each other.
Featured channels may be chosen by YouTube content creators: if your friend has a channel and you want to support it, you can put it on your "Featured Channels" tab.
Related channels are created by YouTube's recommender system.

\item 
We repeat step (3), iteratively collecting another hop of featured/recommended channels from those obtained in (3).
\end{enumerate}

\noindent
The annotation process done here followed the same instructions as the one explained in detail for data collection step \textbf{(c)}.
Steps (2)---(4), were done by a co-author with more than $50$ hours of watch-time of the communities of interest. 
Notice that, in steps (2)---(4), we are not labeling the channels, but creating a pool of channels to be further inspected and labeled in subsequent steps.
The complete list of seeds obtained from (1) and of keywords used in (2) may be found in Appendix~A.
A clear distinction between featured and recommended channels may be found in Appendix~B.

\xhdrNoPeriod{(b)}
For each channel, we collect the number of subscribers and views, and for their videos, all the comments and captions. 
Video and channel recommendations were collected separately using custom-made crawlers. 
We collected multiple "rounds" of recommendations, $22$ for channel recommendations and $19$ for video recommendations. 
Each "round" consists of collecting all recommended channels (on the channel web page) and all recommended videos (on the video web page).
To circumvent possible location bias in the data we collected we used VPNs from 7 different locations: 3 in the USA, 2 in Canada, 1 in Switzerland and 1 in Brazil.
Moreover, channels were always visited in random order, to prevent any biases from arising from session-based recommendations.
As we extensively discuss throughout the paper, this does not include personalization, as we do not log in into any account.

\xhdrNoPeriod{(c)} Channel labeling was done in multiple steps. 
All channels are either seeds (\textit{Type 1}) or obtained through YouTube's recommendation/search engine (\textit{Types 2 and 3)}. 
Notice that \textit{Type 1} channels were assigned labels at the time of their collection. 
For the others, we had 2 of the authors annotate them carefully. 
They both had significant experience with the communities being studied, and were given the following instructions:

\begin{quote}
\small
Carefully inspect each one of the channels in this table,
taking a look at the most popular videos, and watching, altogether, at least 5 minutes of content from that channel. 
Then you should decide if the channel belongs to the Alt-right, the Alt-lite, the Intellectual Dark Web (I.D.W.), or whether you think it doesn't fit any of the communities.
To get a grasp on who belongs to the I.D.W., read ~\cite{weissOpinionMeetRenegades2018}, and check out the website with some of the alleged members of the group ~\cite{anonymousIntellectualDarkWeb}. 
Yet, we ask you to consider the label holistically, including channels that have content from these creators and with a similar spirit to also belong in this category.
To distinguish between the Alt-right and the Alt-lite, read ~\cite{AltRightAlt2019ADL} and \cite{marantzAltRightBrandingWar2017}. 
It is important to stress the difference between civic nationalism and racial nationalism in that case.
Please consider the Alt-right label only to the most extreme content.
You are encouraged to search on the internet for the name of the content creator to help you make your decision.
\end{quote}

\noindent
The annotation process lasted for 3 weeks.
In case they disagreed, they had to discuss the cases individually until a conclusion was reached. 
Interanotator agreement was of $75.57\%$ ($95\%$ CI $[67.5,82.5]$). 
We ended up with $85$ I.D.W., $112$ Alt-lite and $84$ Alt-right channels.

\begin{table}[b]
\small
\centering
\caption{Overview of our dataset.}
\begin{tabular}{lr|lr}
\toprule
Channels &  \numchannels & Video Recs rounds &  19 \\
Videos &  \numvideos & Video Recs &  2,474,044 \\
Comments &  \numcomments & Channel Recs Rounds &  22 \\
Commenting users &  5,980,709 & Channel Recs &  14,283 \\
\bottomrule
\end{tabular}
\label{tab:dataset_ov}
\end{table}

\xhdr{Media}  We also collect popular media channels. 
These were obtained from the \textit{mediabiasfactcheck.com}~\cite{MediaBiasFactMediaBias}. 
For each media source of the categories on the website (\textit{Left, Left-Center, Center, Right-Center, Right}) we search for its name on YouTube and consider it if there is a match in the first page of results~\cite{MediaBiasFactMediaBias}. Some of the channels were not considered because they had too many videos ($15,000+$) and we were not able to retrieve them all (which is important, because our analyses are temporal).
In total, we collect $68$ channels that way.
We use these media channels as a sanity check to capture general trends among more mainstream YouTube channels.

We summarize the dataset collected in the Tab.~\ref{tab:dataset_ov}. 
Data collection was performed during the 19-30th of May 2019, and the collection of the recommendations between May-July 2019. 

\begin{figure*}[t]
\center{\includegraphics[width=\textwidth]
{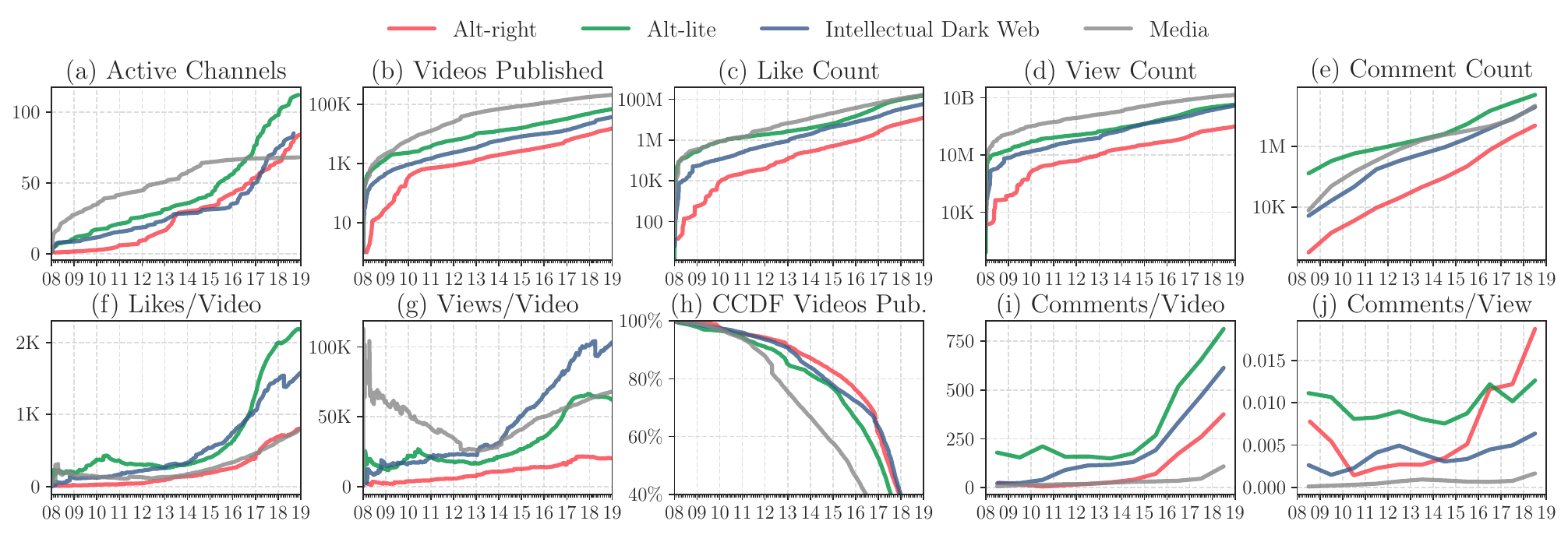}}
\caption{
On the top row figures (a)---(e), for each community and media channels, we have the cumulative number of active channels (that posted at least one video), of videos published, of likes, views and of comments. 
In the bottom row, we have engagement metrics (accumulated over time), (figures (f), (g), (i) and (j)) and the CCDF of videos published, zoomed in the range $[40\%, 100\%]$ on the y-axis (figure (h)). Notice that for comments, we know only the year when they were published, and thus the CDFs granularity is coarser (years rather than seconds).
The raw numbers of views, likes, videos published and more are shown in Appendix~C}
\label{fig:cdfs}
\end{figure*}

\section{The Rise of Contrarians}
\label{sec:004}
We present an overview of the channels in the communities of interest, and show results about their growth in the last years, setting the stage to more in-depth analyses in later sections.
Tab.~\ref{tab:tablechannels} shows the $16$ most viewed YouTubers for each of the communities and for the media channels, and Figure~\ref{fig:cdfs} shows information on the number of videos published, channels created, likes, views, and comments per year, as well as several engagement metrics.

\xhdr{Recent rise in activity} 
Figs.~\ref{fig:cdfs}(a)---(e) show the rise in channel creation, video publishing, likes, views, and comments in the last decade. 
The four latter are growing exponentially for all the communities of interest and for the media channels. 
Noticeably, the rise in the number of active channels is much more recent for the communities of interest than for media channels, as shown in Fig.~\ref{fig:cdfs}(a). 
In mid 2015, for example, $66$ out of the $68$ of the media channels were active (posted their first video), while less than $50\%$ of the Alt-lite, Alt-right and I.D.W. channels had done so. 
This growth in the communities of interest during 2015 may also be noted in Fig.~\ref{fig:cdfs}(i), which shows the CDF of number comments per videos, and
can also be seen between early 2014 and late 2016 in Figs.~\ref{fig:cdfs}(f)---(g), which show the number of likes and views per video, respectively. 
Notice that the number of likes and views is obtained during data collection, and thus, it might be that older videos from those channels became popular later.
Altogether, our data corroborates with the narrative that these communities gained traction in (and fortified) Donald Trump's campaign during the 2016 presidential elections~\cite{campaignsAltLiteBloggersConservative, How2015FueledBuzzFeedNews}.
 
\xhdr{Engagement} 
A key difference between the communities of interest and the media channels is the level of engagement with the videos, as portrayed by the number of likes per video, comments per video and comments per view, shown in Figs.~\ref{fig:cdfs} (f), (i), and (j), respectively. 
For all these metrics, the communities of interest have more engagement than the media channels:
Although media channels have more views per video, as shown in  Figs.~\ref{fig:cdfs}(g), these views are less often converted into likes and comments. 
Notably, Alt-right channels have, since 2017, become the ones with the highest number of comments per view, with nearly $1$ comment per $5$ views by 2018.

\xhdr{Dormant Alt-right Channels} Although by 2013, approximately the same number of channels of all three communities had become active ($\sim 30$), as it can be seen in Fig.~\ref{fig:cdfs}(a), the number of videos they published by the Alt-right was low before 2016.
This can be seen in the CCDF in Fig.~\ref{fig:cdfs}(h): while media and Alt-lite channels had published nearly $40\%$ of their content, the Alt-right had published a bit more than $20\%$. 
This is not because the most popular channels did not yet exist: 4 out of the 5 current top Alt-right channels (accumulating approximately $150$M views) had already been created by 2013. 
Moreover, it is noteworthy that many of the channels now dedicated to Alt-right content have initial videos related to other subjects. 
Take for example the channel ``The Golden One'', number $5$ in Tab.~\ref{tab:tablechannels}. 
Most of the initial videos in the channel are about working out or video-games, with politics related videos becoming increasingly occurring.
The growth in engagement metrics such as likes per video and comments per video of the Alt-right succeeds that of the I.D.W. and of the Alt-lite, resonating with the narrative that the rise of Alt-Lite and I.D.W. channels created fertile grounds for individuals with fringe ideas to prosper~\cite{nagleKillAllNormies2017, lewisAlternativeInfluenceBroadcasting2018}. 

Although our data-driven analysis sheds light on existing narratives on the communities of interest, it is still impossible to determine, from these simple CDFs, whether there is a radicalization pipeline. 
To do so, in the following two sections, we dig deeper into the relationship between these communities looking closely at the users who commented on them.

\begin{figure*}[t]
\center{\includegraphics[width=\linewidth]
{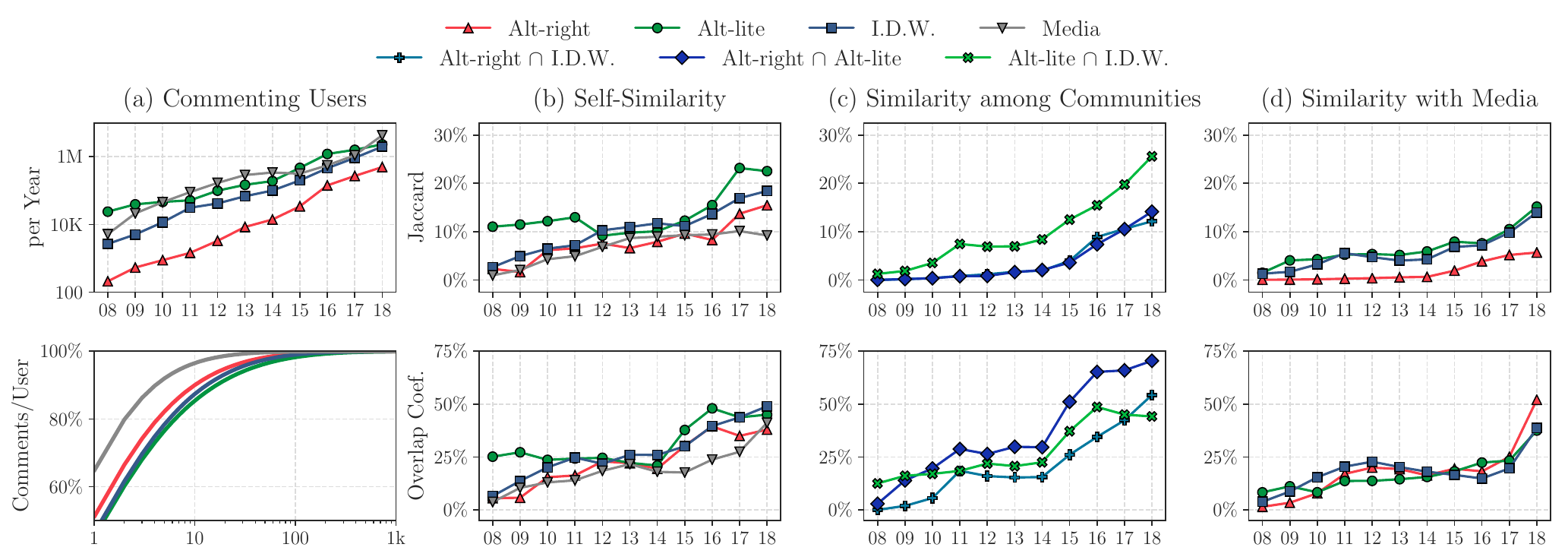}}
\caption{
In \textit{(a)}, the number of unique commenting users per year in the top plot and the CDF of comments per user for each one of the communities in the bottom plot. 
In \textit{(b)}---\textit{(d)} we show two similarity metrics (Jaccard and Overlap Coefficient) for different pairs of sets of commenting users across the years. 
In \textit{(b)} these pairs are the sets of users of each community in subsequent years.
In \textit{(c)} these pairs are the sets of users of each one of the communities of interest.
In \textit{(d)} these pairs are the sets of users of the communities compared with the users who commented in media channels.}
\label{fig:inter_com}
\end{figure*}
\section{User Intersection}
\label{sec:005}
We begin our in-depth analysis of users who commented on the channels of interest by analysing the \emph{intersection} between the users in different channels and communities.
In that context, we use two set similarity metrics:
the Jaccard Similarity $\frac{|A \cap B|}{|A \cup B|}$;
and the Overlap Coefficient $\frac{|A \cap B|}{\min(|A|,|B|)}$.
Notice that the overlap coefficient is particularly useful to compare communities of different sizes. 
For example, a small subset  of a large set may yield low Jaccard Similarity, but will necessarily yield an Overlap Coefficient of 1.

Column (a) of Fig.~\ref{fig:inter_com} characterizes commenting users. 
The top plot shows the absolute number of commenting users per year, while the bottom one shows the CDF of the number of comments per user per community. 
It is interesting to compare these plots with that of Fig.~\ref{fig:cdfs}(e), as we can see that the communities of interest have many more highly active commenters. 
This supports the hypothesis that users who consume content in the communities of interest are more "engaged" than those who consume the content from the media channels. 
Notice also that, although the Alt-right commenters have, on average, fewer comments than those in Alt-lite or the I.D.W., the community is much younger (as discussed in Sec.~\ref{sec:004}), and thus it is hard to tell whether their users are less engaged.

In columns (b)---(d) of Fig. \ref{fig:inter_com} we consider the intersection between the commenting users of the I.D.W., the Alt-lite, the Alt-right and media channels. 
The top figure for each column shows the Jaccard Similarity and the bottom one shows the Overlap Coefficient. 

Column (b) in Fig.~\ref{fig:inter_com} shows the similarity measures for a community with itself a year before (which here we name self-similarity). 
We find that the retention of users among the three communities is growing with time for both metrics. 
However, for media channels, we find that the Jaccard similarity is plateauing since $2014$ and that the overlap coefficient only recently started to grow, perhaps due to the sharp increase in commenting users since $2015$. 
Commenting users from the communities of interest seem to go back more often than those in media channels.

Column (c) in Fig.~\ref{fig:inter_com} shows the pairwise similarity between the three communities.
Notably, in 2018, the Jaccard Similarity between the Alt-lite and the I.D.W. reached almost $30\%$, which is more than the self-similarity between the two communities. 
Moreover, the Overlap Coefficient of the Alt-right with the Alt-lite and the I.D.W is high: reaching around $50\%$ in 2018.   
This means around half of the users who commented in Alt-right channels commented in the other communities.

Lastly, column (d) in Fig.~\ref{fig:inter_com} shows the similarity of the three communities with the media channels. 
We have that the Jaccard similarity between the I.D.W. and the Alt-lite and the media channels is not so different from the similarity between these communities and the Alt-right.
This is a subtle finding.
On one hand, it means that individuals in these communities make up a significant portion of the massive media channels we collected, which gather billions of views.
These communities do not exist in a vacuum but are part of the existing online information environment.
On the other, it shows that the Alt-right, a group of channels with order of magnitudes fewer views, subscribers and comments, are actually \textit{on par} with these large channels.
Inspecting the Overlap Coefficient, however, we get a different view: there we have that the communities overlap more with themselves than with the media channels, particularly since $2015$. 
However, in $2018$, there is a sharp growth in the similarity with media channels. 
A hypothesis for this is that, as these channels grew more popular (as previously discussed in Sec.~\ref{sec:004}, they became more mainstream).

These analyses take us one step further in understanding the communities being studied.
We again see that their users are more engaged, and, notably, find that the I.D.W, the Alt-lite, and the Alt-right increasingly share the same commenting user base.

\begin{figure*}[ht]
\center{\includegraphics[width=\linewidth]
{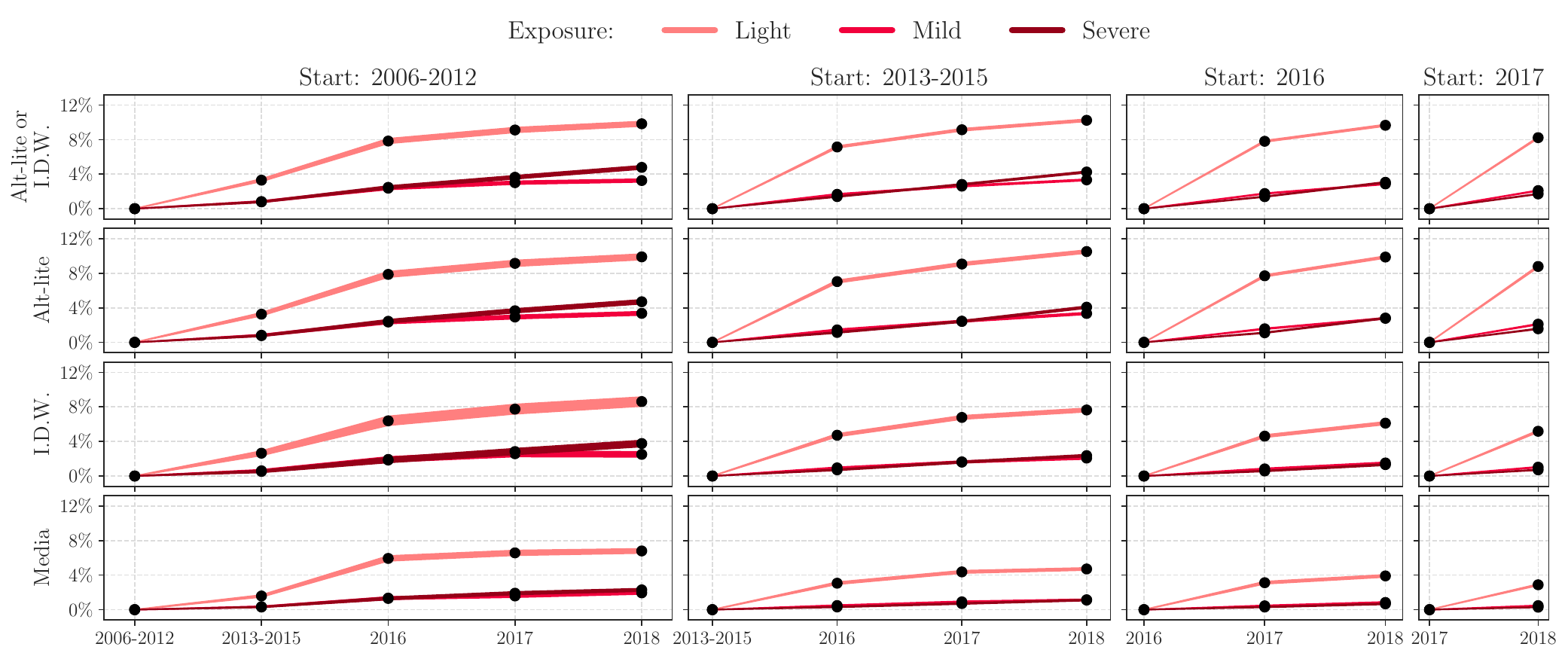}}
\caption{We show how users migrate towards Alt-right content. 
For users who consumed only videos in the communities indicated by the labels in the rows (Alt-lite or I.D.W., Alt-lite, I.D.W., and Media), we show the chance that they go on to consume Alt-right content. 
We consider three levels of exposure: light (commented in 1 to 2 Alt-right videos), mild (3 to 5) and severe (6+). 
Each column tracks users on a different starting date. 
Initially, their exposure rates are 0 (as they did not consume any Alt-right content). 
As time passes, we show the exposure rates in the y-axis, for each of the years, in the x-axis. Line widths represent $95\%$ confidence intervals.}
\label{fig:prospective}
\end{figure*}
\section{User Migration}
\label{sec:006}
In the previous section, we showed that the commenting user bases among the I.D.W., the Alt-lite, and the Alt-right are increasingly similar ---and the effect is stronger than for media channels. 
This indicates that there is a growing percentage of users consuming extreme (Alt-right) content on YouTube \textit{while also} consuming content from other milder communities (Alt-lite/I.D.W.). 
Yet, it does not, \textit{per se}, indicate that there is a radicalization pipeline on the website. 
It could be, for example, that new users who join the website go on to consume content from all three communities.
To better address this question, we find users who \textit{did not comment} in Alt-right content in a given year and track their subsequent activity. 
Notice that we do not have the user's entire activity history, and thus, we track their activity only in the channels whose videos we collected. 

For four time brackets $[(2006-2012),(2013-2015), (2016), (2017)]$ we track four sets of users: 
those who only commented on videos of the Alt-lite or the I.D.W, 
those who did so only for videos on the Alt-lite, 
those who did so only for videos on the I.D.W.,
and those who commented only on videos of the media channels.
Then, for subsequent years, we track the same users. 
Notice that when users are tracked for one year they are not eligible for selection in upcoming years. 
We consider these users to be exposed if they commented on $1$-$2$ (light), $3$-$5$ (mild) or $6+$ (severe) Alt-right videos.

The results for this analysis are shown in Fig.~\ref{fig:prospective}.
We show the percentage of users we managed to track that were exposed. 
The number of users tracked and exposed at each step may be found in Appendix~C.
Consider, for example, users who on $2006-2012$ commented only on  I.D.W. or Alt-lite content (227,945 users), as shown in the subplot in the first column and the first row.
By $2018$, around $10\%$ were lightly exposed, and roughly $4\%$ severely or mildly so --- which amounts to approximately 9K users in total.
From the ones who in $2017$ commented only on Alt-lite or I.D.W. videos (1,251,674 users), as shown in the last column of the first row, approximately  $12\%$ of them were exposed --- more than 60K users altogether. 

We also find that media channels present lower exposure rates, as can be seen in the last row of the figure.
The difference is particularly large for the last three time brackets.
Less than $1\%$ of users in media channels were mildly or severely exposed, against $3\%$ to $4\%$ for Alt-lite or I.D.W. users, and roughly $4\%$ were lightly exposed versus approximately $8\%$ for Alt-lite or I.D.W users.

When teasing apart users that commented only on Alt-lite or only on I.D.W. content, we find that, not only users who commented \emph{only} on I.D.W. get less exposed, but increasingly less so. The same applies to the media channels. 
For example, the exposure rates of users who watched only Alt-lite (second row) or only I.D.W. (third row) content are much more similar for those tracked in $2006-2012$ (first column) than for those tracked in $2017$ (last column). 
For users who were tracked in $2006-2012$, around $15\%$ were exposed in both scenarios, while for those tracked in $2017$, this difference grew farther apart ($\sim12\%$ Alt-lite vs. $\sim6\%$ I.D.W.). 

\begin{figure*}[ht]
\center{\includegraphics[width=\linewidth]
{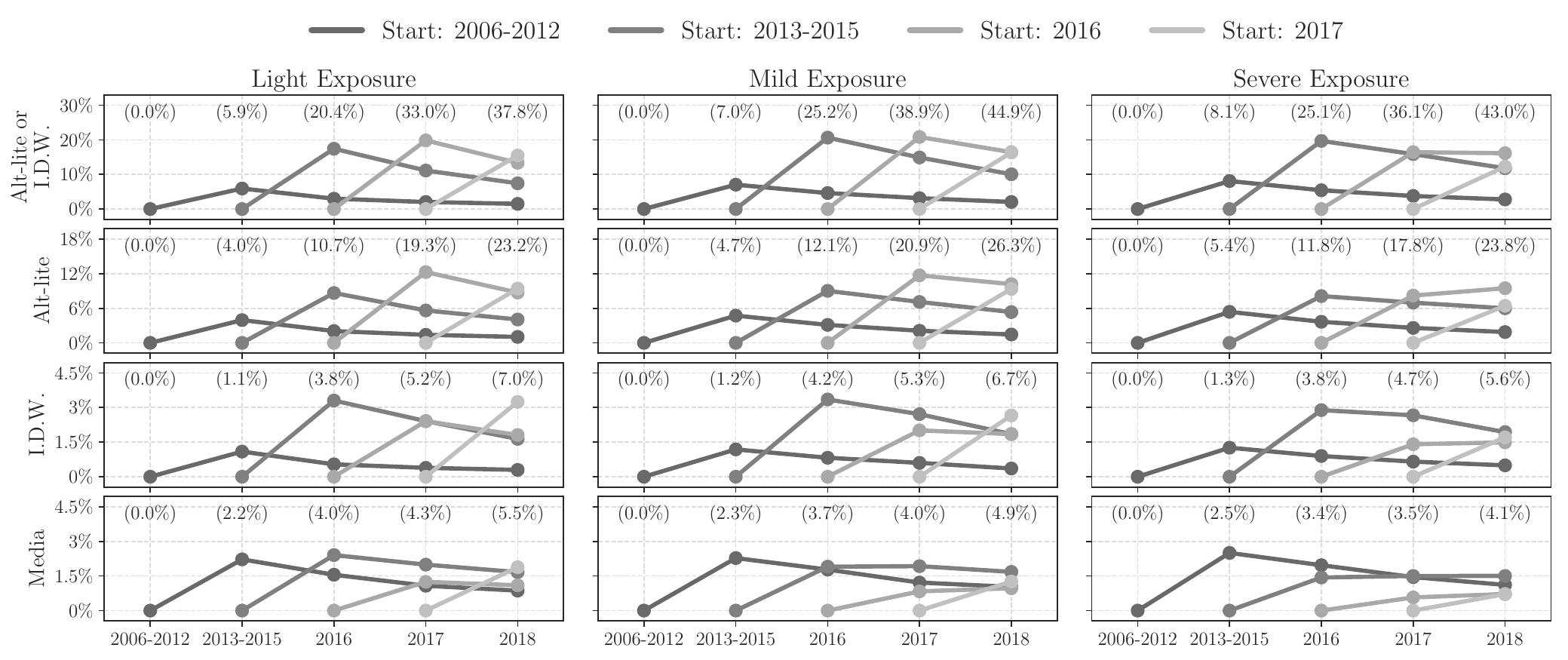}}
\caption{We show how expressive the tracked users are in terms of the Alt-right user base.
Each row shows a different condition for tracking users and each column shows a different level of exposure.
Each line corresponds to users tracked at a different starting date (in the x-axis), and the y-axis shows the percentage of the total Alt-right commenting users they went to become (notice that all lines begin at 0, because initially they did not consume any Alt-right content).}
\label{fig:prospective_p}
\end{figure*}

The previous study suggests that the pipeline effect does exist, and that indeed, users systematically go from milder communities to the Alt-right. 
However, it does not give insight into how expressive the effect is in terms of what part of the Alt-right user base has gone through it. 
We address this question by tracking users exactly as we did before, and then analyzing what percentage of exposed users at each year can be traced back to users who initially watched content from other communities. 
In other terms, for each year we calculate, of the users who are exposed (i.e. who watched Alt-right videos), which percentage belongs to each one of the sets of tracked users we just described. 

The results for this analysis are shown in Fig.~\ref{fig:prospective_p}.
We find that these users are a considerable fraction of the Alt-right commenting audience. 
In 2018, for all kinds of exposure, roughly $40\%$ of commenting users can be traced back from cohorts of users that commented only on Alt-lite or I.D.W. videos in the past.
This can be seen in the first row of the plot.
Moreover, we can observe that, consistently, users who consumed Alt-lite or I.D.W. content in a given year, go on to become a significant fraction of the Alt-right user base in the following year.
This number is much more expressive than the number of users which came from media channels --- in the last row --- which never surpasses $6\%$ for any level of exposure.

Looking at the second and third row of Fig.~\ref{fig:prospective_p},  we find a substantial difference between the I.D.W. and the Alt-lite. 
Whereas in Sec.~\ref{sec:005} we find that the intersection between them both and the Alt-right are  similar, here we see that users who initially commented \textit{only} on I.D.W. channels constitute a much less significant percentage of the Alt-right consumer base in upcoming years. 
For all levels of exposure, at all times, the number of exposed users that can be traced back to commenting exclusively on I.D.W. channels is around 3 times lower.
So, while in $2018$, $23.3\%$ of users who were lightly exposed can be traced back to users who commented on  Alt-lite channels in previous years, only $7.6\%$ can be traced back to I.D.W. channels.
Overall, in both analyses, users who consumed only I.D.W. channel seem to behave more similarly to the users in the media channels.
Yet, as we see in Sec.~\ref{sec:005}, the intersection between the Alt-lite and the I.D.W. is increasing with time, which means this population is becoming less significant.

The experiments performed show that, not only the commenting user bases are becoming increasingly similar (as shown in Sec.~\ref{sec:005}), but that, systematically, users who commented only on I.D.W. or Alt-lite content go on to comment on Alt-right channels.
This phenomenon is significant both in terms of the percentage of the users tracked --- as in Fig.~\ref{fig:prospective} --- and in terms of the total Alt-right commenting user base --- as in Fig.~\ref{fig:prospective_p}. We present the raw numbers associated with these figures in Appendix~D.

\begin{figure*}[t]
\begin{minipage}[b]{\linewidth}
\center{\includegraphics[width=\linewidth]{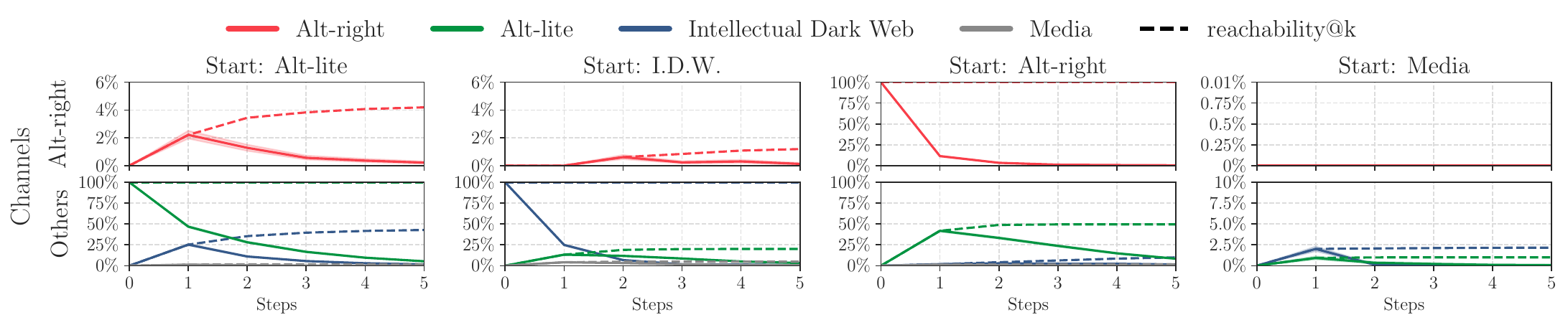}}
\subcaption{\vspace{-0.2\baselineskip}}
\end{minipage}
\begin{minipage}[b]{\linewidth}
\center{\includegraphics[width=\linewidth]{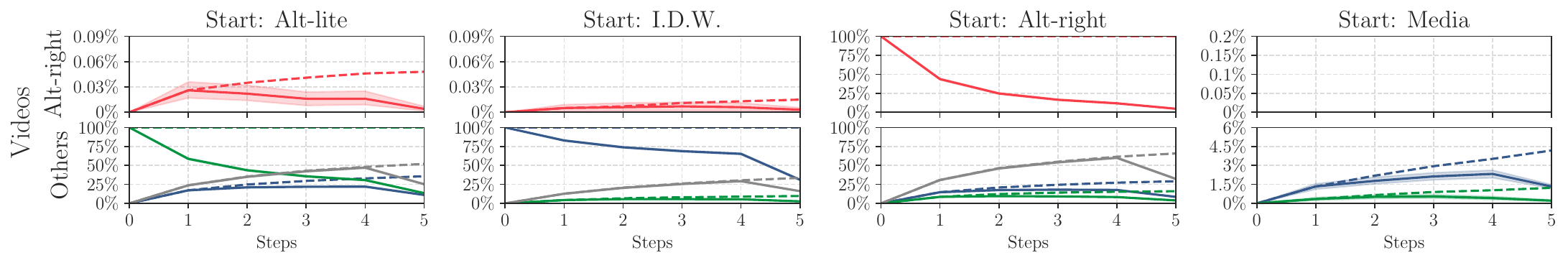}}
\subcaption{\vspace{-0.2\baselineskip}}
\end{minipage}%
\vspace{-0.65\baselineskip}
\caption{We show the results for the simulation of random walks  for channels (a) and videos (b).
We show two metrics, as described in the text, the probability of the walker being in a given community at each step (solid line) and the reachability at each step for a given community (dashed line).
The different columns portray different starting rules for the initial node in the simulations.
Error bands are $95\%$ confidence intervals.
}
\label{fig:rec_graphs}
\end{figure*}

\section{The Recommendation Algorithm}
\label{sec:007}
\begin{table}[b]
\centering
\footnotesize
\caption{Percentage of edges in-between communities in the recommendation graphs (normalized per weight). Video recommendations are in bold. Rows indicate the source of edges columns indicate their destination.}
\begin{tabular}{lrrrrr}
\toprule
\textbf{Src|Dst} & I.D.W. & Alt-lite &  Alt-right & Media & Other \\
\midrule                                                                 
I.D.W.    & 52.78/\textbf{19.03} & 22.88/\textbf{1.57}  &  0/\textbf{0.03} & 3.12/\textbf{3.03}  &  21.23/\textbf{76.35} \\     
Alt-L  & 13.69/\textbf{2.46}  & 55.15/\textbf{12.70} &  3.38/\textbf{0.13} & 2.82/\textbf{3.24}  &  24.96/\textbf{81.47} \\     
Alt-R & 25.73/\textbf{1.89}  & 42.94/\textbf{1.15}  & 25.73/\textbf{8.55} & 1.35/\textbf{3.38}  &  21.08/\textbf{85.03} \\     
Media   &  4.94/\textbf{0.31}  &  4.36/\textbf{0.08}  & 0/\textbf{0}        & 28.78/\textbf{14.84} &  61.92/\textbf{84.77} \\     
\bottomrule                                                              
\end{tabular}

\label{tab:chan_rec}
\end{table}

In this section, we inspect the impact of YouTube's recommendation algorithm.
Unfortunately, we have only a snapshot of the recommender system which does not take into account personalization.
Thus, it is hard to reach significant conclusions on what was the role of the recommender system in the radicalization process we depicted in Sec.~\ref{sec:006}.
Yet, we argue that analyzing these data is relevant, for it is a blueprint of how the influence of the recommender system may be measured, and because it allows us to understand how the recommender system is behaving for our scenario.

We perform our analysis in a recommendation graph, built using the data collected. 
The graph is built as follows: for each channel, we join together all recommendations obtained in all rounds of data collection. 
Each channel is a node, and edges between nodes indicate recommendations from a channel to another (for both video and channel recommendations).
Notice that, in case there was a recommendation towards a channel or a video we are not aware of, we add an edge to a special sink node we name ``Other''.
Each edge is weighted proportionally to the number of times that recommendation appeared in the data collection, and weights are normalized so that outgoing edges of each node sum up to $1$.

The percentage of edges between communities (normalized by their weight) is shown in Tab.~\ref{tab:chan_rec} for channel and video recommendations. 
For channel recommendations, we have that media channels are recommended scarcely by the communities of interest.
In fact, there are more edges flowing \textit{out} of media channels towards Alt-lite/I.D.W. channels than the other way around.
Alt-lite and I.D.W. channels recommend channels from the same community around $50\%$ of the time, and recommend each other around $14\%$ (Alt-L to I.D.W.) and $23\%$ (I.D.W. to Alt-L) of the time. 
Alt-right channels are only recommended by Alt-lite channels ($3.08\%$). 
For video recommendations, there is a high prevalence of recommendation to videos we were not able to track (more than $75\%$ of outgoing edges from all communities pointed towards the ``Other'' node). 
We also find that media channels are more often recommended in this setting ( $\sim 3\%$ for all communities), while the Alt-lite and the I.D.W. recommend each other roughly $2\%$ of the time.
Lastly, Alt-right videos are not significantly recommended here.

Given these graphs, we experiment with random walks. 
The random walker begins in a random node, chosen with chance proportional to the number of subscribers in each channel. 
Then, the random walker randomly navigates the graph for $5$ steps, choosing edges at random with probabilities proportional to their weights. 
We store the random walks and calculate two metrics:
\textit{(i)} the probability of it being in a channel from each of the communities, that is, the probability that there is a channel of a given community in the $k$-th step.
\textit{(ii)} the reachability of each of the communities at step $k$.
That is, at step $k$, the percentage of times that the random walker has found a node of a given community.
We run the simulation 10K times for scenarios where the initial node is restricted to one of the three communities or the media channels.

Importantly, we consider a small difference in the experimental set-up for each of the graphs. 
In the channel recommendation graph, we allow the random walker to choose the ``Other'' node. 
When this happens the walk stops, thus at each step there is a probability this walk is interrupted by this --- or by the fact that there are no recommended channels.
In the channel recommendation graph, as the number of edges to the ``Other'' node is too high, we do not allow the random walker to go towards it. 
Notice that the scenario for the channels is more realistic, and we give more weight to the conclusions drawn there.
The two aforementioned metrics, at each step, given different starting conditions, are shown in Fig.~\ref{fig:rec_graphs}, for channel and video recommendations.

For channel recommendations, we have that the reachability@$5$ of Alt-right channels is of approximately $4\%$ for the simulations starting from Alt-lite $1.5\%$ for I.D.W. channels.
Moreover, starting from an I.D.W. channel, users have approximately $10\%$ of chance of being in an Alt-lite channel at the next step, and in $5$ steps, there is $25\%$ of chance that the user has found at least one Alt-lite channel.
Starting from the media channels, reachability@$5$ of I.D.W. channels is of $2.5\%$, and of slightly less than $1\%$ for Alt-lite channels. These can be seen on the bottom row of Fig.~\ref{fig:rec_graphs} (a).

For video recommendations, reaching Alt-right channels from other communities is less likely. 
From the Alt-lite, reachability@$5$ is of around $0.05\%$.
Going from the I.D.W. to the Alt-lite is more difficult: the reachability@$5$ is roughly $7\%$.
More relevant, though, starting from media channels, the reachability@$5$ of I.D.W. and Alt-lite channels is of around $4.5\%$ and $1.5\%$ respectively.
It is worth recalling that this experiment is less realistic than the former, as here we ignore the possibility of the random walker being in a video we are not aware of.

Overall, we find that, in the channel recommender system, it is easy to navigate from the I.D.W. to the Alt-lite (and vice-versa), and it is possible to find Alt-right channels. 
From the Alt-lite we follow the recommender system $5$ times, approximately $1$ out of each $25$ times we will have spotted an Alt-right channel (as seen in Fig.~\ref{fig:rec_graphs} (a)).
In the video recommender system, Alt-right channels are less recommended, but finding Alt-lite channels from the I.D.W. and I.D.W. channels from the large media channels in the media group is also feasible.
Considering the sheer amount of views the channels in the Alt-lite, the I.D.W. and the Alt-lite, these percentages, although low, may result in a very significant number of views towards fringe content.
This process may also be amplified when taking personalization into account.
Notice that we depict the two graphs in which we performed our experiments in Appendix~E.

\section{Discussion}
\label{sec:008}
We performed a through analysis of three YouTube communities --- the I.D.W., the Alt-lite, and the Alt-right --- inspecting a large dataset with millions of comments and recommendations from thousands of videos. 
In this section, we discuss how the insights of our analyses shed light into our research questions. 
We also talk about the limitations and potential implications of this work.

\xhdrNoPeriod{RQ1. How have these channels grown on YouTube in the last decade?}
The three communities studied sky-rocketed in terms of views, likes, videos published and comments, particularly, since $2015$, coinciding with the presidential election of that year, as shown in Sec.~\ref{sec:004}. 
However, this seems to be the case not only for these communities, but also for the larger channels in the media group.
A key difference between the communities and media channels lies in the engagement of their users.
The number of comments per view seems to be particularly high for extreme content (Sec.~\ref{sec:004}), and users in all three communities are more assiduous commentators than in the media channels (Sec.~\ref{sec:005}).

\xhdrNoPeriod{RQ2. To which extent do users systematically gravitate towards more extreme content?}
We find that the commenting user bases for the three communities are increasingly similar (Sec.~\ref{sec:005}), and, considering Alt-right channels as a proxy for extreme content, that a significant amount of commenting users systematically migrates from commenting exclusively on milder content to commenting on more extreme content (Sec.~\ref{sec:007}).
We argue that this finding provides significant evidence that there has been, and there continues to be, user radicalization on YouTube, and our analyses of the activity of these communities (Sec.~\ref{sec:004}) is consistent with the theory that more extreme content ``piggybacked'' on the surge in popularity of I.D.W. and Alt-lite content~\cite{nagleKillAllNormies2017}.
We show that this migration phenomenon is not only consistent throughout the years, but also that it is significant in its absolute quantity.
Noticeably, the findings related to this research question make the implicit assumption that commenting users are a good enough proxy for radicalization, and that comments in YouTube channels are supportive of the videos they are associated with.
We established the validity of these assumptions as follows.
First, the sheer number of comments and high prevalence of comments per views in Alt-right videos suggest that commenting users are a population worth studying, especially when in Sec.~\ref{sec:004} we found that Alt-right channels have a very high percentage of comments per view.
Secondly, during the three week annotation period, it was noted that the number of opposing comments is rather small, as we found by
manually checking $900$ randomly selected comments  ($300$ for each community of interest), finding that only 5 could be interpreted as criticisms to the videos they were associated with.
Moreover, we note that the proportion of likes for the communities of interest is higher for the communities of interest ($>91\%$ mean, $>96\%$ median) than for the media channels ($85\%$ mean, $93\%$ median), which suggests the people interacting with the three  communities agree with their videos.

\xhdrNoPeriod{RQ3. Do algorithmic recommendations steer users towards more extreme content?}
Our simulations suggest that YouTube's recommendation algorithms frequently suggest Alt-lite and I.D.W. content. 
From these two communities, it is possible to find Alt-right content from recommended channels, but not from recommended videos.
Noticeably, our analysis has several shortcomings which do not allow us to make bold claims about this research question. 
Firstly, we are able to look only at a tiny fraction of actual recommendations --- it could very well be that Alt-right content was being more widely promoted in the past.
Secondly, our analysis does not take into account personalization, which could reveal a completely different picture.
Still, even without personalization, we were still able to find a path in which users could find extreme content from large media channels.

\xhdrNoPeriod{Limitations and future work.}
Our work resonates with the narrative that there is a radicalization pipeline~\cite{rooseMakingYouTubeRadical2019, tufekciOpinionYouTubeGreat2018}. 
Indeed, we manage to measure traces of user radicalization using commenting users.
Although we argue this is strong evidence for the existence of radicalization pathways on YouTube, our work provides little insight on \textit{why} these radicalization pipelines exist.
% They could possibly be explained by the way the YouTube platform works (allowing for content creators to produce niche content).
% Also, radicalization towards fringe ideologies has happened in times previous to social media, and may be influenced by other societal factors.
% Trying to tease apart the influence of these different factors (platform, algorithm, society)
Elucidating the causes of radicalization
is an important direction to better understand user radicalization and the influence of social media in our lives.
Moreover, in this paper we focused exclusively on basic statistics (likes, views and comments) and on the trajectory of users, be they inferred through comments or simulated in the recommendation graphs.
Another interesting direction would be to trace the evolution of the \textit{speech} of content creators and commenting users throughout the years, to study what are the narratives that arose and how their tone has changed. 

\xhdr{Acknowledgements} We gratefully acknowledge support from the Brazilian agencies CNPq, Capes and Fapemig, from the projects Atmosphere, INCT-Cyber and MASWEB, from a Google Research Award for Latin America (Manoel Horta Ribeiro). We thank Jeremy (Jimmy) Blackburn for helpful discussions.
\newpage

\bibliographystyle{ACM-Reference-Format}
\bibliography{000_main} 

\newpage
\appendix

\section{Data Collection}
\label{app:201}
\begin{figure}[t]
\center{\includegraphics[width=0.57\linewidth]
{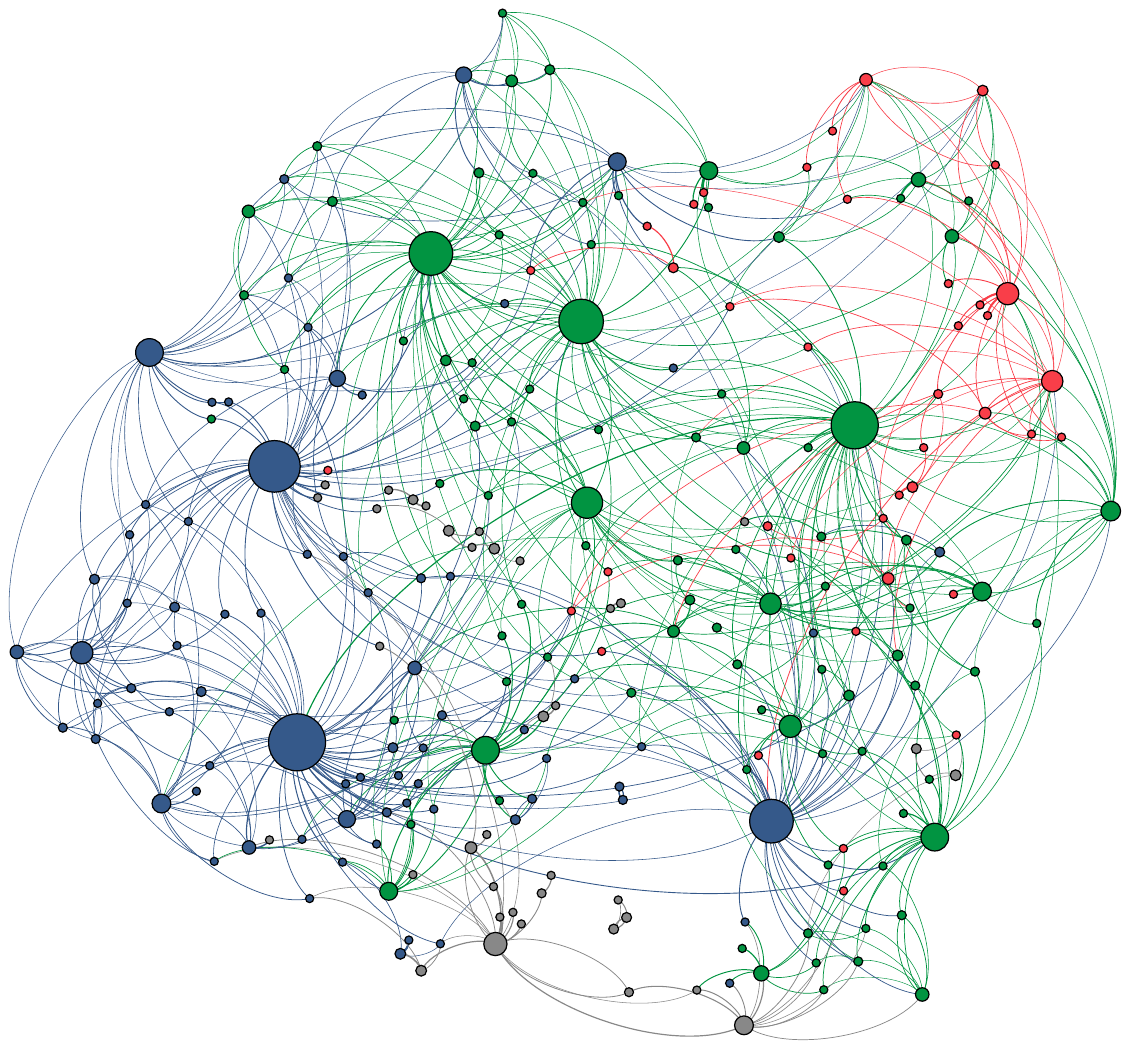}}
\caption{Recommendation graph of YouTube channels. }
\label{fig:channel_g}
\end{figure}

\begin{figure}[t]
\center{\includegraphics[width=0.75\linewidth]
{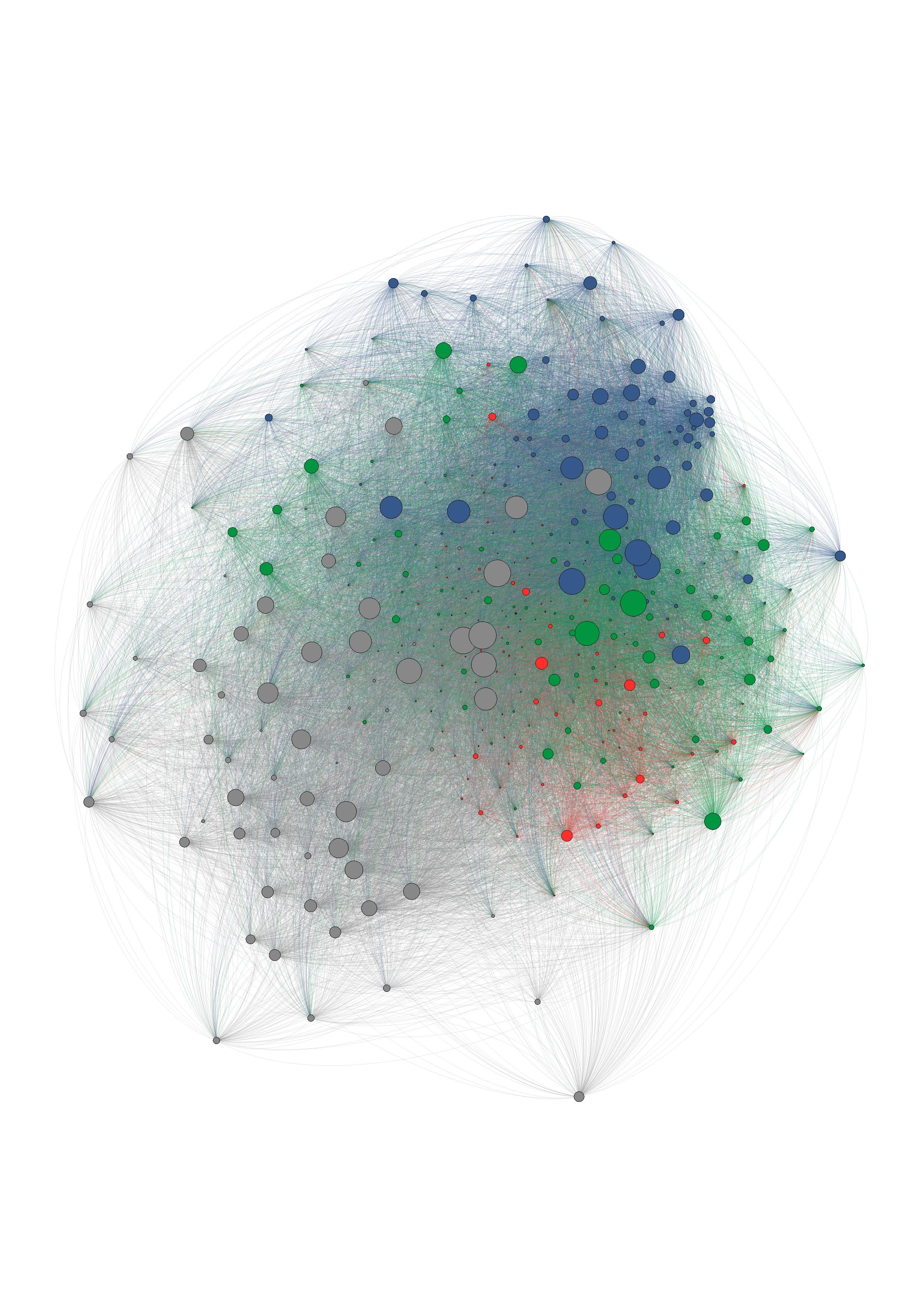}}
\caption{Recommendation graph of YouTube videos. 
Colors for communities are the same as those in the paper.}
\label{fig:video_g}
\end{figure}

\begin{figure}[t]
\center{\includegraphics[width=\linewidth]
{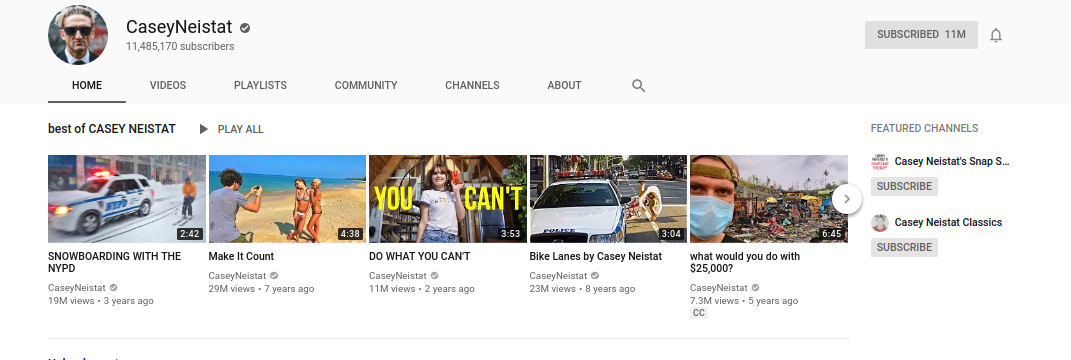}}
\caption{Example of YouTube channel with featured channel on the side.}
\label{fig:featured}
\end{figure}

We give some details in the data collection process.
 Tab.~\ref{tab:tableall1} and Tab.~\ref{tab:tableall2} show for channels labeled as Alt-right, Alt-lite and I.D.W., their communities and data collection steps. Tab ~\ref{tab:tablecontrol} shows all media channels we obtained.
Fig.~\ref{fig:example_data_collec} highlights what was collected on YouTube. Below, we enumerate the keywords employed to search for channels of each of the communities:

\xhdrNoPeriod{For the I.D.W.}
\textit{Stephen Hicks, Camille Paglia, Carl Benjamin, Elon Musk, Akira the Don, Nicholas Christakis, Claire Lehmann, Matt Christiansen, Steven Pinker, Rebel Wisdom, Tim Pool, Quillette, Jonathan Haidt, Peter Thiel, Lindsay Shepherd, James Damore}

\xhdrNoPeriod{For the Alt-lite}
\textit{Brittany Pettibone, Jack Posobiec, Gavin McInnes, Kyle Chapman, Kyle Prescott, Lucian Wintrich, Mike Cernovich, Milo Yiannopoulos, Stefan Molyneaux, Vee, Blonde in the Belly of the Beast, Paul Joseph Watson, Styxhenxenhammer666, Rebel Media, Lauren Chen, Computing Forever, Andy Warski, Owen Benjamin, Steven Crowder
}

\xhdrNoPeriod{For the Alt-right}
\textit{Evola, Evropa, The Jewish Question, White Genocide, Mass immigration, Andrew Anglin, weev, Andy Nowicki, Augustus Invictus, Christopher Cantwell, Collin Liddell, Daniel J. Kleve, Daniel Friberg, Dillon Irizarry, Greg Johnson, Jared Taylor, Jason Kessler, Jason Reza Jorjani, Johnny Monoxide, Lana Lokteff, Matt Forney, Matthew Heimbach, Matthew Parrott, Mike Enoch, Nathan Damigo, Pax Dickinson, Richard Spencer, Tara McCarthy, Vox Day, Baked Alaska}

\section{Featured vs Recommended}
\label{app:202}
\begin{figure*}[ht]
\center{\includegraphics[width=0.95\textwidth]
{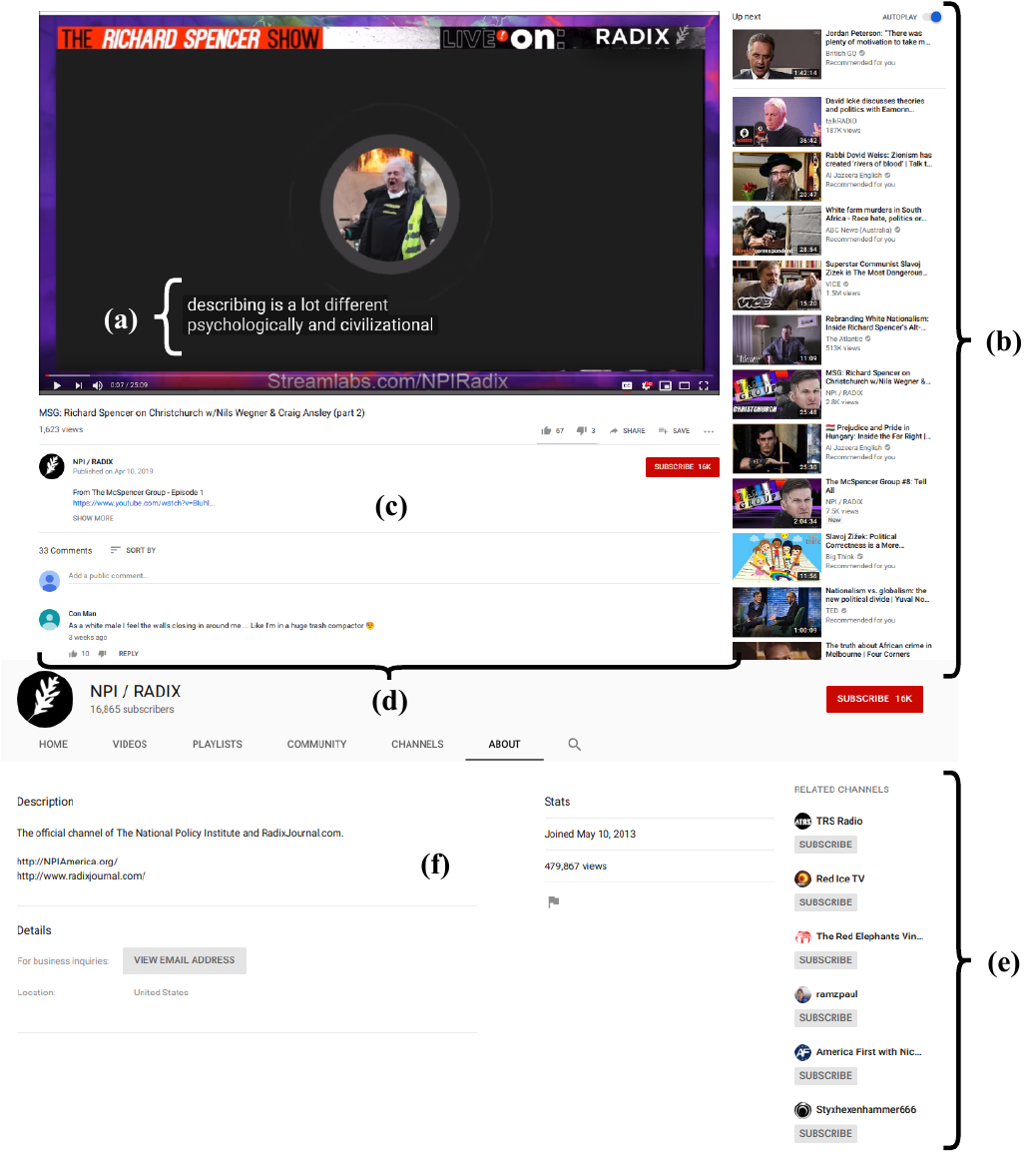}}
\caption{Overview of the elements we collected: (a) video captions, when available, (b) video recommendations, (c) video description and metadata, (d) comments, (e) channel recommendations, and (f) video metadata.}
\label{fig:example_data_collec}
\end{figure*}

We illustrate the difference between featured and recommended channel.
In Fig.~\ref{fig:featured} you may see an example of featured channels, these are chosen by the channel owner.
In Fig.~\ref{fig:example_data_collec}, letter (e) shows related channels, these are recommendations made by YouTube.

\section{Likes, Videos, Views, Comments}
\label{app:203}
Tab.~\ref{tab:104_table_views} shows, for the three communities, the number of likes, views, videos and commenting users across the years.

\begin{table*}[t]
\caption{For all categories, we list the number of likes, views, videos and commenting users across the years.}
\begin{tabular}{llrrrr}
\toprule
Category &  Year &  Like Count &   View Count &  Video Count &  Comment Count \\
\midrule
Alt-lite              &  2008 &    272639 &   18145720 &         1392 &         129130 \\
 &  2009 &    585060 &   32993863 &          929 &         197934 \\
 &  2010 &    503744 &   30519109 &         1498 &         248432 \\
 &  2011 &    527221 &   28400257 &         2344 &         236961 \\
 &  2012 &    805166 &   61779929 &         4142 &         360688 \\
 &  2013 &   1237131 &  101136564 &         2476 &         531614 \\
 &  2014 &   2574586 &  158822748 &         3319 &         824757 \\
 &  2015 &   8227303 &  398745164 &         7299 &        2787437 \\
 &  2016 &  27277364 & 1001985084 &         9442 &        8983525 \\
 &  2017 &  55014745 & 1393845365 &        15858 &       12322105 \\ 
 &  2018 &  54723719 & 1092143577 &        20681 &       21341673 \\ \midrule
Alt-right             &  2008 &       559 &      73159 &           29 &            332 \\
 &  2009 &      8389 &    1236895 &          313 &           1135 \\
 &  2010 &     14124 &    1897892 &          363 &           2136 \\
 &  2011 &     15992 &    1778120 &          174 &           6076 \\
 &  2012 &     75092 &    4925868 &          434 &          10452 \\
 &  2013 &    160494 &   11219639 &          654 &          25825 \\
 &  2014 &    233381 &   12718956 &          725 &          47032 \\
 &  2015 &    434925 &   17148672 &          958 &         127319 \\
 &  2016 &   1411778 &   44177307 &         2334 &         529821 \\
 &  2017 &   4253888 &  108482909 &         3548 &        1262549 \\
 &  2018 &   5773031 &  106455102 &         5843 &        2725573 \\ \midrule
Media &  2008 &    348137 &  128986765 &         2115 &           7932 \\
 &  2009 &    511468 &  196992273 &         3939 &          41492 \\
 &  2010 &    573299 &  203399250 &         6531 &          94379 \\
 &  2011 &   1824078 &  350120542 &        12748 &         200385 \\
 &  2012 &   3432239 &  454969357 &        25716 &         447302 \\
 &  2013 &   5238196 &  716009326 &        18135 &         756691 \\
 &  2014 &   9217725 & 1538251895 &        18836 &         814124 \\
 &  2015 &  16569182 & 2015671151 &        24168 &         830655 \\
 &  2016 &  27807514 & 2481994316 &        30119 &        1317648 \\
 &  2017 &  46467022 & 3102590498 &        35678 &        2931209 \\
 &  2018 &  54106314 & 2997876294 &        30951 &       13667470 \\ \midrule
I.D.W. &  2008 &     54185 &    7034287 &          447 &           5263 \\
 &  2009 &     61340 &    8661426 &          488 &          11249 \\
 &  2010 &    135205 &   15457288 &          549 &          29932 \\
 &  2011 &    269021 &   22797055 &          884 &         129453 \\
 &  2012 &    365241 &   23918023 &         1067 &         154322 \\
 &  2013 &   1085086 &  109350887 &         2520 &         226833 \\
 &  2014 &   2451712 &  230853763 &         2304 &         367374 \\
 &  2015 &   3297678 &  261930446 &         2053 &         858816 \\
 &  2016 &   6616069 &  447000398 &         3291 &        2056423 \\
 &  2017 &  18820727 & 1135173733 &         8789 &        4291180 \\
 &  2018 &  25625709 & 1575685392 &        14842 &       11013421 \\
\bottomrule
\end{tabular}
\label{tab:104_table_views}
\end{table*}

\section{User Trajectories}
\label{app:204}
Tab.~\ref{tab:106_trajectories} shows the absolute numbers of users tracked and infected (at all levels, as mentioned in Sec.~\ref{sec:006}. 
It also shows what percentage of the total number of users who watched Alt-right the number of users infected was. 
Additionally, In Tab.~\ref{tab:107_trajectories}, we show the trajectories from the Alt-right to the other two communities and media channel (that is, we repeat the exact same procedure tracking users from the Alt-right and checking if they commented in the other communities).
We find that users from the Alt-right diffuse in similar propotions to the other communities and the media channels.

\begin{table*}[t]
\caption{We show absolute numbers for users infected and tracked in Sec.~\ref{sec:006}, as well as what percentage of the total number of users who watched Alt-right the number of users infected was.}
\footnotesize
\begin{tabular}{lllrrrr}
\toprule
Category & Start & Year &   \# Users Infected &  \# Users Tracked & \% of Users Alt-right & \%  Tracked and Infected \\
\midrule
Alt-lite &2006-2012&2006-2012&               0 &         170301 &           0\% &                  0\% \\
      &   &2013-2016&            2132 &          43872 &         146\% &                 4.86\% \\
      &   &2016&            3426 &          27045 &          8.83\% &                12.67\% \\
      &   &2017&            4558 &          28944 &           6.1\% &                15.75\% \\
      &   &2018&            6186 &          34436 &          4.37\% &                17.96\% \\
      &2013-2016&2013-2016&               0 &         414353 &           0\% &                  0\% \\
      &   &2016&           12287 &         127591 &          25.8\% &                 9.63\% \\
      &   &2017&           16345 &         117181 &         19.71\% &                13.95\% \\
      &   &2018&           22753 &         126815 &         15.39\% &                17.94\% \\
      &2016&2016&               0 &         718464 &           0\% &                  0\% \\
      &   &2017&           31290 &         301252 &         32.19\% &                10.39\% \\
      &   &2018&           45005 &         290816 &          28.4\% &                15.48\% \\
      &2017&2017&               0 &         777106 &           0\% &                  0\% \\
      &   &2018&           44017 &         352938 &         25.22\% &                12.47\% \\ \midrule
Alt-lite or I.D.W. &2006-2012&2006-2012&               0 &         227945 &           0\% &                  0\% \\
      &   &2013-2015&            3192 &          64874 &         215\% &                 4.92\% \\
      &   &2016&            5016 &          39527 &         132\% &                12.69\% \\
      &   &2017&            6645 &          42140 &          8.94\% &                15.77\% \\
      &   &2018&            8895 &          49748 &          6.29\% &                17.88\% \\
      &2013-2015&2013-2015&               0 &         694155 &           0\% &                  0\% \\
      &   &2016&           25848 &         252962 &         57.75\% &                10.22\% \\
      &   &2017&           33532 &         230172 &         41.91\% &                14.57\% \\
      &   &2018&           42629 &         239116 &         29.27\% &                17.83\% \\
      &2016&2016&               0 &        1040872 &           0\% &                  0\% \\
      &   &2017&           52610 &         480309 &         579\% &                10.95\% \\
      &   &2018&           70905 &         454870 &         45.91\% &                15.59\% \\
      &2017&2017&               0 &        1251674 &           0\% &                  0\% \\
      &   &2018&           74534 &         619501 &          44.1\% &                123\% \\ \midrule
Media &2006-2012&2006-2012&               0 &         248214 &           0\% &                  0\% \\
      &   &2013-2015&            1136 &          50724 &           7\% &                 2.24\% \\
      &   &2016&            2331 &          27168 &           5.3\% &                 8.58\% \\
      &   &2017&            3123 &          30991 &          3.75\% &                108\% \\
      &   &2018&            4629 &          41913 &          31\% &                114\% \\
      &2013-2015&2013-2015&               0 &         637338 &           0\% &                  0\% \\
      &   &2016&            3146 &          81489 &          5.75\% &                 3.86\% \\
      &   &2017&            5159 &          86127 &          5.42\% &                 5.99\% \\
      &   &2018&            8123 &         116469 &          4.85\% &                 6.97\% \\
      &2016&2016&               0 &         365614 &           0\% &                  0\% \\
      &   &2017&            2929 &          75512 &          2.65\% &                 3.88\% \\
      &   &2018&            5000 &          92281 &          2.79\% &                 5.42\% \\
      &2017&2017&               0 &         696297 &           0\% &                  0\% \\
      &   &2018&            7809 &         214600 &          3.86\% &                 3.64\% \\ \midrule
I.D.W. &2006-2012&2006-2012&               0 &          47914 &           0\% &                  0\% \\
      &   &2013-2015&             565 &          14948 &          3.54\% &                 3.78\% \\
      &   &2016&             889 &           8745 &          2.26\% &                10.17\% \\
      &   &2017&            1240 &           9424 &          1.64\% &                13.16\% \\
      &   &2018&            1686 &          11322 &          1.15\% &                14.89\% \\
      &2013-2015&2013-2015&               0 &         212122 &           0\% &                  0\% \\
      &   &2016&            4634 &          72573 &          9.54\% &                 6.39\% \\
      &   &2017&            6742 &          67199 &          7.79\% &                103\% \\
      &   &2018&            8543 &          70677 &          5.43\% &                129\% \\
      &2016&2016&               0 &         232159 &           0\% &                  0\% \\
      &   &2017&            5942 &          98640 &          5.83\% &                 62\% \\
      &   &2018&            8711 &          97288 &          5.15\% &                 8.95\% \\
      &2017&2017&               0 &         420116 &           0\% &                  0\% \\
      &   &2018&           14268 &         206053 &          7.61\% &                 6.92\% \\
\bottomrule
\end{tabular}
\label{tab:106_trajectories}
\end{table*}

\begin{table*}[t]
\caption{We show absolute numbers for users infected and tracked in Sec.~\ref{sec:006}, as well as what percentage of the total number of users who watched Alt-right the number of users infected was.}
\footnotesize
\begin{tabular}{lllrrrr}
\toprule
Category & Start & Year &   \# Users Infected &  \# Users Tracked & \% of Users Alt-right & \%  Tracked and Infected \\
\midrule
Alt-right to I.D.W. & 2006-2012& 2006-2012&               2006-2012&           3276 &           0\% &                  0\% \\
          &   & 2013-2015&             283 &            997 &          0.24\% &                28.39\% \\
          &   & 2016&             216 &            569 &          0.13\% &                37.96\% \\
          &   & 2017&             290 &            646 &           0.1\% &                44.89\% \\
          &   & 2018&             391 &            741 &          07\% &                52.77\% \\
          & 2013-2015& 2013-2015&               2006-2012&          21578 &           0\% &                  0\% \\
          &   & 2016&            1421 &           6384 &          0.72\% &                22.26\% \\
          &   & 2017&            1821 &           5691 &          0.59\% &                 32\% \\
          &   & 2018&            2565 &           5922 &          0.43\% &                43.31\% \\
          & 2016& 2016&               2006-2012&          41385 &           0\% &                  0\% \\
          &   & 2017&            4142 &          15752 &          1.17\% &                 26.3\% \\
          &   & 2018&            5511 &          15426 &          0.87\% &                35.73\% \\
          & 2017& 2017&               2006-2012&          69241 &           0\% &                  0\% \\
          &   & 2018&            9024 &          29987 &          1.31\% &                309\% \\ \midrule
Alt-right to Alt-lite & 2006-2012& 2006-2012&               2006-2012&           3276 &           0\% &                  0\% \\
          &   & 2013-2015&             465 &            997 &          0.29\% &                46.64\% \\
          &   & 2016&             399 &            569 &          0.12\% &                70.12\% \\
          &   & 2017&             410 &            646 &           0.1\% &                63.47\% \\
          &   & 2018&             407 &            741 &          06\% &                54.93\% \\
          & 2013-2015& 2013-2015&               2006-2012&          21578 &           0\% &                  0\% \\
          &   & 2016&            3047 &           6384 &          0.87\% &                47.73\% \\
          &   & 2017&            2962 &           5691 &          0.64\% &                525\% \\
          &   & 2018&            2884 &           5922 &          0.43\% &                 48.7\% \\
          & 2016& 2016&               2006-2012&          41385 &           0\% &                  0\% \\
          &   & 2017&            7696 &          15752 &          1.49\% &                48.86\% \\
          &   & 2018&            7089 &          15426 &          0.98\% &                45.95\% \\
          & 2017& 2017&               2006-2012&          69241 &           0\% &                  0\% \\
          &   & 2018&           13435 &          29987 &          1.79\% &                 44.8\% \\ \midrule
Alt-right to Media & 2006-2012& 2006-2012&               0.0 &           3276.0 &           0.0\% &                  0.0\% \\
          &   & 2013-2015&             407.0 &            997.0 &          0.21\% &                40.82\% \\
          &   & 2016&             152.0 &            569.0 &          0.12\% &                26.71\% \\
          &   & 2017&             225.0 &            646.0 &          0.09\% &                34.83\% \\
          &   & 2018&             406.0 &            741.0 &          0.05\% &                54.79\% \\
          & 2013-2015& 2013-2015&               0.0 &          21578.0 &           0.0\% &                  0.0\% \\
          &   & 2016&            1043.0 &           6384.0 &           0.6\% &                16.34\% \\
          &   & 2017&            1504.0 &           5691.0 &          0.49\% &                26.43\% \\
          &   & 2018&            2872.0 &           5922.0 &          0.29\% &                 48.5\% \\
          & 2016& 2016&               0.0 &          41385.0 &           0.0\% &                  0.0\% \\
          &   & 2017&            4787.0 &          15752.0 &          1.41\% &                30.39\% \\
          &   & 2018&            8639.0 &          15426.0 &          0.84\% &                 56.0\% \\
          & 2017& 2017&               0.0 &          69241.0 &           0.0\% &                  0.0\% \\
          &   & 2018&           16269.0 &          29987.0 &          1.46\% &                54.25\% \\
\bottomrule
\end{tabular}
\label{tab:107_trajectories}
\end{table*}

\section{Recommendation Graphs}
\label{app:205}
In Figs. ~\ref{fig:channel_g} and ~\ref{fig:video_g} we show the recommendation graphs used for the experiment in Section~\ref{sec:007}.

\begin{table*}[t]
\caption{For the three communities, we list all the websites analysed in this paper (part 1). \textbf{Edit (21-Oct-2021)}: Upon request on the channel owner’s behalf, we have removed the channel ‘theglassblindspot’, which was incorrectly labeled, from the table. Since the channel is small, removing it has no noticeable impact on the results presented in this paper.}
\small
\begin{tabular}{llllrll}
{} & Alt-right channels & Step & Alt-lite channels & Step & I.D.W. channels & Step \\
\midrule
0   &                                AltRight.com &              1 &  America First with Nicholas J Fuentes &              1 &                          Ben Shapiro &                          1 \\
1   &                              AmRen Podcasts &              1 &                            Andy Warski &              1 &                       Bret Weinstein &                          1 \\
2   &                                 AmRenVideos &              1 &       Blonde in the Belly of the Beast &              1 &                             Gad Saad &                          1 \\
3   &            Ayla Stewart Wife With A Purpose &              1 &                     Brittany Pettibone &              1 &                            JRE Clips &                          1 \\
4   &                              Baked Alaska 2 &              1 &                      Computing Forever &              1 &              Jordan B Peterson Clips &                          1 \\
5   &                         Black Pigeon Speaks &              1 &                          Gavin McInnes &              1 &                 JordanPetersonVideos &                          1 \\
6   &                                Bre Faucheux &              1 &                           Laura Loomer &              1 &                     Lindsay Shepherd &                          1 \\
7   &                           CounterCurrentsTV &              1 &                            Lauren Chen &              1 &                    Matt Christiansen &                          1 \\
8   &                                  Darkstream &              1 &                        Lauren Southern &              1 &                        Owen Benjamin &                          1 \\
9   &                               Faith J Goldy &              1 &                                   MILO &              1 &                  Owen Benjamin Clips &                          1 \\
10  &                                James Allsup &              1 &                         Mike Cernovich &              1 &                          PowerfulJRE &                          1 \\
11  &                               Jason Kessler &              1 &                     Nick Fuentes Clips &              1 &                         Rebel Wisdom &                          1 \\
12  &                       Jean-François Gariépy &              1 &                            No Bullshit &              1 &                           Sam Harris &                          1 \\
13  &                             Johnny Monoxide &              1 &                          No Bullshit 2 &              1 &                     SargonofAkkad100 &                          1 \\
14  &                                     MW Live &              1 &                     Paul Joseph Watson &              1 &                     The Rubin Report &                          1 \\
15  &                                 Matt Forney &              1 &                           Rebel Canada &              1 &                       joerogandotnet &                          1 \\
16  &                              MillennialWoes &              1 &                             Rebel Edge &              1 &                                 1791 &                          2 \\
17  &                                 NPI / Radix &              1 &                            Rebel Media &              1 &                     American Justice &                          2 \\
18  &                                  Red Ice TV &              1 &                        Stefan Molyneux &              1 &  Atheist Foundation of Australia Inc &                          2 \\
19  &                                Staying Woke &              1 &                          StevenCrowder &              1 &                     AynRandInstitute &                          2 \\
20  &                              The Golden One &              1 &                     Styxhexenhammer666 &              1 &                Ben Shapiro Thug Life &                          2 \\
21  &                      The Reality Calls Show &              1 &                           The Thinkery &              1 &                     Benjamin A Boyce &                          2 \\
22  &                 Traditionalist Worker Party &              1 &                                    Vee &              1 &                    Brother Nathanael &                          2 \\
23  &                                 Voxiversity &              1 &                              6oodfella &              2 &                               CISAus &                          2 \\
24  &                         augustussolinvictus &              1 &                               A1Cvenom &              2 &                       Clash of Ideas &                          2 \\
25  &                              iambakedalaska &              1 &                       AIU-Resurrection &              2 &      Conversations with Bill Kristol &                          2 \\
26  &                                   Alt Right &              2 &                         AltRight Truth &              2 &                               Crysta &                          2 \\
27  &                            Alt-Right Tankie &              2 &                      AustralianNeoCon1 &              2 &                    Desi-Rae Thinking &                          2 \\
28  &                              American Pride &              2 &                                BlazeTV &              2 &               Douglas Murray Archive &                          2 \\
29  &                            American Pride 2 &              2 &                        Brave New World &              2 &                       Enlightainment &                          2 \\
30  &                                ArktosOnline &              2 &                             Bull Brand &              2 &                      Essential Truth &                          2 \\
31  &  Augustus Invictus for United States Senate &              2 &                          Carpe Donktum &              2 &                       Freedom Speaks &                          2 \\
32  &                           AustralianRealist &              2 &                   Christopher Anderson &              2 &                           Glenn Beck &                          2 \\
33  &                              Be Open MInded &              2 &                           Daily Caller &              2 &                            Gravitahn &                          2 \\
34  &                     BigCatKayla Livestreams &              2 &                       DailyCallerVideo &              2 &                          Informative &                          2 \\
35  &                              Charles Zeiger &              2 &                              DailyKenn &              2 &          Jordan Peterson Fan Channel &                          2 \\
36  &                               Corpus Mentis &              2 &                         Dinesh D'Souza &              2 &                           Liberty us &                          2 \\
37  &                        Dismantle The Matrix &              2 &                      DoctorRandomercam &              2 &                                   MG &                          2 \\
38  &                              Dissident View &              2 &                     Domination Station &              2 &               Maximilien Robespierre &                          2 \\
39  &                                   Engländer &              2 &                    Harrison Hill Smith &              2 &                     MeaningofLife.tv &                          2 \\
40  &                                 Jan Kerkoff &              2 &                             Jacob Wohl &              2 &                           Mike Nayna &                          2 \\
41  &                                Mark Collett &              2 &                              Kelly Day &              2 &                       Motte \& Bailey &                          2 \\
42  &                               Matthew North &              2 &                           Leo Stratton &              2 &                              MrAndsn &                          2 \\
43  &               Nacionalista Blanco del SoCal &              2 &                   Liberty Machine News &              2 &               Notes For Space Cadets &                          2 \\
44  &                   Nationalist Media Network &              2 &                              Luke Ford &              2 &                             Pangburn &                          2 \\
45  &                              No White Guilt &              2 &                  Luke Ford Livestreams &              2 &                   PhilosophyInsights &                          2 \\
46  &                            Patrick Slattery &              2 &                Make Cringe Great Again &              2 &              Pragmatic Entertainment &                          2 \\
47  &                                  Real McGoy &              2 &                             News2Share &              2 &                             ReasonTV &                          2 \\
48  &                                Revcon Media &              2 &                       On The Offensive &              2 &                         Savage Facts &                          2 \\
49  &                             Stand Up Europe &              2 &                        Oppressed Media &              2 &                      The Daily Truth &                          2 \\
\bottomrule
\end{tabular}
\label{tab:tableall1}
\end{table*}

\begin{table*}[t]
\caption{For the three communities, we list all the websites analysed in this paper (part 2).}
\small
\begin{tabular}{llllrll}
{} & Alt-right channels & Step & Alt-lite channels & Step & I.D.W. channels & Step \\
\midrule
50  &                              Steve Trueblue &              2 &                     Revenge Of The Cis &              2 &                 The Free Speech Club &                          2 \\
51  &                  The Alternative Hypothesis &              2 &                          RobinHoodUKIP &              2 &              The Heritage Foundation &                          2 \\
52  &                          The Great Dolemite &              2 &                     SJW CRINGE MACHINE &              2 &                    The New Criterion &                          2 \\
53  &                The James Delingpole Channel &              2 &                             SJWCentral &              2 &                The Pondering Primate &                          2 \\
54  &                              The Last Stand &              2 &                             Semiogogue &              2 &               The Unplugged Observer &                          2 \\
55  &                           The Rational Rise &              2 &                   Social Justice Fails &              2 &                      TheArchangel911 &                          2 \\
56  &                           TheArmenianNation &              2 &                       The Fallen State &              2 &                      TheAtlasSociety &                          2 \\
57  &                              This is Europa &              2 &                  &  &                         Transliminal &                          2 \\
58  &                          ThuleanPerspective &              2 &                      The Hateful Gaels &              2 &                  Trigger Happy Media &                          2 \\
59  &                Traditionalist Youth Network &              2 &                         The Iconoclast &              2 &                              VikNand &                          2 \\
60  &                     Truth Against The World &              2 &                       TheSchillingShow &              2 &                     Washington Watch &                          2 \\
61  &                          WhiteRabbitRadioTV &              2 &  Tipping Point With Liz Wheeler on OAN &              2 &                          WisdomTalks &                          2 \\
62  &                                andy nowicki &              2 &                         Tommy Robinson &              2 &                                YAFTV &                          2 \\
63  &                                   eliharman &              2 &                          Tree Of Logic &              2 &                             ZIEeICoZ &                          2 \\
64  &                              jackburton2009 &              2 &                    UNITE AMERICA FIRST &              2 &                        ZeroFox Given &                          2 \\
65  &                               nightmarefuel &              2 &                            Western Man &              2 &                        battleofideas &                          2 \\
66  &                             14 Sacred Words &              3 &                              Zach Hing &              2 &                     bloggingheads.tv &                          2 \\
67  &                              Awakened Saxon &              3 &                              grapjas60 &              2 &                              bmdavll &                          2 \\
68  &                             Borzoi Boskovic &              3 &                           hOrnsticles3 &              2 &                       successcouncil &                          2 \\
69  &                                  Danny 1488 &              3 &                               ramzpaul &              2 &                           tmcleanful &                          2 \\
70  &                          InvincibleNumanist &              3 &                               theovonk &              2 &                        wikileaksplus &                          2 \\
71  &                               LaughingMan0X &              3 &                     theturningpointusa &              2 &                       xUnlimitedMagz &                          2 \\
72  &                                Laura Towler &              3 &                              thkelly67 &              2 &                               ybrook &                          2 \\
73  &                         LibertarianRealist2 &              3 &                 Actual Justice Warrior &              3 &                     AgatanFoundation &                          3 \\
74  &                           Little Revolution &              3 &                        AllNationsParty &              3 &                Bite-sized Philosophy &                          3 \\
75  &                                Marie Cachet &              3 &                       Alt Hype Streams &              3 &                        CoolHardLogic &                          3 \\
76  &                                    Morrakiu &              3 &                          Aydin Paladin &              3 &                        Davie Addison &                          3 \\
77  &                    NeatoBurrito Productions &              3 &                        Beacom Of Light &              3 &                        Dose of Truth &                          3 \\
78  &                                NewEuropeANP &              3 &                                Bearing &              3 &                     DronetekPolitics &                          3 \\
79  &                                  OnlineWipe &              3 &                          Count Dankula &              3 &          Galactic Bubble Productions &                          3 \\
80  &                             Oswald Spengler &              3 &                            Dangerfield &              3 &                     HowTheWorldWorks &                          3 \\
81  &                          Prince of Zimbabwe &              3 &                                Demirep &              3 &                     ManOfAllCreation &                          3 \\
82  &                                   Serp Kerp &              3 &                       Dr. Steve Turley &              3 &                     PragerUniversity &                          3 \\
83  &                                   TRS Radio &              3 &                           IRmep Stream &              3 &                          Rekt Idiots &                          3 \\
84  &                               The Leftovers &              3 &                          Jericho Green &              3 &                         Sinatra\_Says &                          3 \\
85  &                                    The Lion &              3 &                              John Ward &              3 &                   Sorting Myself Out &                          3 \\
86  &              The Revolutionary Conservative &              3 &                      JustInformed Talk &              3 &               The Andrew Klavan Show &                          3 \\
87  &                              VertigoPolitix &              3 &                        Liberty Hangout &              3 &                       The Daily Wire &                          3 \\
88  &                                      &         &                            MR. OBVIOUS &              3 &              The Heartland Institute &                          3 \\
89  &                                      &         &                               MarkDice &              3 &           The Propertarian Institute &                          3 \\
90  &                                      &         &                         MichelleRempel &              3 &                              Timcast &                          3 \\
91  &                                      &         &                         Mister Metokur &              3 &                               &                     \\
92  &                                      &         &                         NateTalksToYou &              3 &                               &                     \\
93  &                                      &         &                          OneTruth4Life &              3 &                               &                     \\
94  &                                      &         &                        ProductiehuisEU &              3 &                               &                     \\
95  &                                      &         &                Reverend Simon Sideways &              3 &                               &                     \\
96  &                                      &         &                        Sanity 4 Sweden &              3 &                               &                     \\
97  &                                      &         &                   Sargon of Akkad Live &              3 &                               &                     \\
98  &                                      &         &                           SkidRowRadio &              3 &                               &                     \\
99  &                                      &         &                     Slightly Offens*ve &              3 &                               &                     \\
100 &                                      &         &                              Tea Clips &              3 &                               &                     \\
101 &                                      &         &                      The Amazing Lucas &              3 &                               &                     \\
102 &                                      &         &                       The Weekly Sweat &              3 &                               &                     \\
103 &                                      &         &                           TheBechtloff &              3 &                               &                     \\
104 &                                      &         &                  TheIncredibleSaltMine &              3 &                               &                     \\
105 &                                      &         &                          Toad McKinley &              3 &                               &                     \\
106 &                                      &         &                   TokenLibertarianGirl &              3 &                               &                     \\
107 &                                      &         &                               Undoomed &              3 &                               &                     \\
108 &                                      &         &     Vincent James of The Red Elephants &              3 &                               &                     \\
109 &                                      &         &                                 ataxin &              3 &                               &                     \\
110 &                                      &         &                          brianoflondon &              3 &                               &                     \\
111 &                                      &         &                                jaydyer &              3 &                               &                     \\
112 &                                      &         &                      libertydollshouse &              3 &                               &                     \\
113 &                                      &         &                             patcondell &              3 &                               &                     \\
\bottomrule
\end{tabular}
\label{tab:tableall2}
\end{table*}

\begin{table*}[t]
\caption{Media channels.}
\small
\begin{tabular}{llllll}
\toprule
{} & Left & Center & Left-Center & Right-Center & Right \\
\midrule
0  &                cosmopolitan &                big think &        (the)atlantic &    forbes &     american enterprise institute \\
1  &               democracy now &  c-span &     business insider & gulf news &   judicial watch \\
2  &                 elite daily &         consumer reports &             cbc news &              learn liberty &  national rifle association (nra) \\
3  &               good magazine &          financial times &             engadget &              new york post &         pj media \\
4  &                 gq magazine &  harvard business review &   feminist frequency &  ntd.tv (new tang dynasty) &  project veritas \\
5  &  huffington post (huffpost) &             investopedia &     glamour magazine &             russia insider &           ron paul liberty report \\
6  &   mashable &                makeuseof &       global citizen &       &              \\
7  & merry jane &             mental floss &          global news &       &              \\
8  &           new york magazine &             military.com &   hollywood reporter &       &              \\
9  & new yorker &  recode &             la times &       &              \\
10 &             people magazine &        relevant magazine &           lifehacker &       &              \\
11 &      slate &            the economist &  new york daily news &       &              \\
12 &     uproxx &       the indian express &        rolling stone &       &              \\
13 &   upworthy &        today i found out &  san francisco globe &       &              \\
14 &                 vanity fair & vocativ &           scoopwhoop &       &              \\
15 &        vox &     world economic forum &            scroll.in &       &              \\
16 &        &     &             sky news &       &              \\
17 &        &     &           techcrunch &       &              \\
18 &        &     &         the guardian &       &              \\
19 &        &     &            the verge &       &              \\
20 &        &     &            vice news &       &              \\
21 &        &     &      washington post &       &              \\
22 &        &     &       wired magazine &       &              \\
23 &        &     &           yahoo news &       &              \\
\bottomrule
\end{tabular}
\label{tab:tablecontrol}
\end{table*}

\end{document}